\documentclass[12pt]{article}
\usepackage{amsmath}
\usepackage{latexsym}
\usepackage{epsfig}
\usepackage{graphics}
\title{{\bf Analytical On-shell Calculation of Higher Order Scattering: Massless Particles}}
\author{Barry R. Holstein\\
Department of Physics-LGRT\\
University of Massachusetts\\
Amherst, MA  01003\\
and\\
Kavli Institute for Theoretical Physics\\
University of California\\
Santa Barbara, CA  93016}
\begin{titlepage}
\begin{document}
\maketitle
\begin{abstract}
We demonstrate that the use of on-shell methods involving calculation of the discontinuity across the t-channel cut associated with the exchange of a pair of massless particles can be used to evaluate loop contributions to the electromagnetic and gravitational scattering of both massive and massless particles.  In the gravitational case the use of factorization permits a straightforward and algebraic calculation of higher order scattering results, which were obtained previously by considerably more arduous Feynman diagram techniques.
\end{abstract}
\end{titlepage}

\section{Introduction}

Many investigations have been reported concerning higher order effects in electromagnetic scattering~\cite{Iwa71},\cite{Spr93},\cite{Fei88},\cite{Ros08}, gravitational scattering~\cite{Don94},\cite{Muz95},\cite{Ham95},\cite{Akh97},\cite{Khr03},\cite{Bje03}\cite{Ros08a} and both~\cite{But06},\cite{Fal08},\cite{Ros08b}.  The goal of such calculations has typically been to seek an effective potential which characterizes these higher order effects.  For both electromagnetic and gravitational interactions, the leading potential is well-known and has the familiar $1/r$ fall-off with distance.  The higher order contributions required by quantum mechanics lead to corrections which are shorter range, from local to polynomial fall-off---$1/r^n$ with $n\ge 2$.  The effective potential is defined to be the Fourier transform of the nonrelativistic scattering amplitude via\footnote{Note that Eq. (\ref{eq:nb}) follows from the Born approximation for the scattering amplitude
\begin{equation}
{\rm Amp}(\boldsymbol{q})=<\boldsymbol{p}_f|\hat{V}|\boldsymbol{p}_i>=\int d^3re^{i\boldsymbol{q}\cdot\boldsymbol{r}}V(r)
\end{equation}
and nonrelativistic amplitudes are defined by dividing the covariant forms by the normalizing factor $4E_AE_B\simeq 4m_Am_B$.}
\begin{equation}
V(r)=-\int{d^3q\over (2\pi)^3}e^{-i\boldsymbol{q}\cdot\boldsymbol{r}}{\rm Amp}(\boldsymbol{q})
\end{equation}
where $\boldsymbol{q}=\boldsymbol{p}_i-\boldsymbol{p}_f$ is the three-momentum transfer.  Thus, for lowest order one-photon or one-graviton exchange, the dominant momentum-transfer dependence arises from the propagator
\begin{equation}
{1\over q^2}\stackrel{NR}{\longrightarrow}{-1\over\boldsymbol{q}^2}+{\cal O}(\boldsymbol{q}^4)
\end{equation}
since the energy transfer $q_0={\cal O}(\boldsymbol{q}^2)$ in the nonrelativistic limit.  The Fourier transform
\begin{equation}
\int{d^3q\over (2\pi)^3}e^{-i\boldsymbol{q}\cdot\boldsymbol{r}}{1\over \boldsymbol{q}^2}={1\over 4\pi|\boldsymbol{r}|}
\end{equation}
then leads to the expected $1/r$ dependence.  By dimensional analysis it is clear that shorter range $1/r^2$ and $1/r^3$ behavior can arise only from {\it nonanalytic}
$1/|\boldsymbol{q}|$ and $\ln \boldsymbol{q}^2$ dependence, which arise from higher order scattering contributions.  Any analytic momentum dependence from polynomial contributions in such diagrams can lead only to short-distance ($\delta^3(\boldsymbol{r})$ and derivatives) effects.  Thus, if we are seeking only the longest range corrections, we need identify only the nonanalytic components of the higher order contributions to the scattering amplitude.

The basic idea behind use of the on-shell method is that the scattering amplitude must satisfy the stricture of unitarity, which requires that
its discontinuity across the right hand cut is given by
\begin{equation}
 {\rm Disc}\, T_{fi}=i(T_{fi}-{T^\dagger}_{fi})=-\sum_n T_{fn}T^\dagger_{ni}\label{eq:mk}
\end{equation}
By requiring that this condition be satisfied, we guarantee that the nonanalytic structure will be maintained, and in a companion paper we showed how this program is carried out in the case of the electromagnetic and gravitational scattering of spinless particles, both of which have mass~\cite{Hol16}.\footnote{Note that in order to make the present paper self contained, we have repeated here some material contained in \cite{Hol16}.}  In the present work we extend this work to review the scattering of two massive particles and extend the calculation to the case wherein one of the scattered particles becomes massless.  Specifically, in  the next section we show how this on-shell procedure is used to evaluate the electromagnetic scattering of two spinless particles, considering both massive and massless possibilities.  Then in section 3, we demonstrate how, using the property of factorization, this technique can be easily generalized to the case of gravitational scattering.  Our conclusions are summarized in a brief closing section.  Three appendices contain some of the calculational details.

\section{On-Shell Method: Electromagnetic Scattering}

We begin with the electromagnetic scattering of two charged spinless particles $A$ and $B$, with mass, charge $m_A,\,e$ and $m_B,\,e$ respectively.  In this case the
intermediate state sum is over two-photon states, and we require the product of Compton annihilation and creation amplitudes---$P+P\rightarrow\gamma+\gamma\rightarrow P'+P'$.  The electromagnetic interaction has the form
\begin{equation}
{\cal L}_{int}=-eA^\mu J_\mu
\end{equation}
where $e=\sqrt{4\pi\alpha}$ is the electric charge, $A^\mu$ is the electromagnetic vector potential, and $J_\mu$ is the electromagnetic current, which at leading order has the matrix element between spin zero particles
\begin{equation}
<p_2|J_\mu|p_1>=(p_1+p_2)_\mu +{\cal O}(q)
\end{equation}
where $q=p_1-p_2$.  The corresponding Compton scattering amplitude is easily found~\cite{Hol14}
\begin{equation}
{\rm Amp}_0^A(\epsilon_1,\epsilon_2)=2e^2\left(\epsilon_1^*\cdot\epsilon_2^*-{\epsilon_1^*\cdot p_1\epsilon_2^*\cdot p_2\over p_1\cdot k_1}-{\epsilon_1^*\cdot p_2\epsilon_2^*\cdot p_1\over p_1\cdot k_2}\right)\,,
\end{equation}
which results from the sum of the three diagrams shown in Figure 1.

Consider now the amplitude
\begin{equation}
{\rm Amp}_2^{em}(q)={1\over 2!}{1\over 4m_Bm_B}\int{d^4\ell\over (2\pi)^4}{d^4\ell'\over (2\pi)^4}(2\pi)^4\delta^4(p_1+p_2-\ell-\ell'){1\over \ell^2{\ell'}^2}{{\cal N}_{em}\over \prod_{i=1}^4 p_i\cdot\ell}\label{eq:xv}
\end{equation}
where ${\cal N}_{em}$ is chosen so that, on-shell\footnote{Note that because of the on-shell condition, we can replace any of the terms $p_i\cdot\ell$ in the denominator by $(\ell-p_i)^2-m^2$ without altering the discontinuity. Only analytic (short distance) pieces of the amplitude are modified.}
\begin{equation}
{{\cal N}_{em}\over \prod_{i=1}^4 p_i\cdot\ell}=\sum_{a,b=1}^2{\rm Amp}_0^A(\epsilon_1^{*a},\epsilon_2^{*b})\left({\rm Amp}_0^{B*}(\epsilon_1^{*a},\epsilon_2^{*b})\right)^*
\end{equation}
Using the Cutkosky rules, we see that this amplitude satisfies the on-shell unitarity relation
\begin{eqnarray}
{\rm Disc}\,{\rm Amp}_2^{em}(q)&=&-{i\over 2!}{e^4\over 8m_Am_B}\int {d^3\ell\over (2\pi)^32\ell_0}{d^3\ell'\over (2\pi)^32\ell'_0}(2\pi)^4\delta^4(p_1+p_2-\ell-\ell'){{\cal N}_{em}\over \prod_{i=1}^4p_i\cdot\ell}\nonumber\\
&=&-i{e^4\over 128\pi m_Am_B}<{{\cal N}_{em}\over \prod_{i=1}^4p_i\cdot\ell}>\label{eq:xz}
\end{eqnarray}
so that ${\rm Amp}_2^{em}(q)$ coincides with the actual scattering amplitude up to {\it analytic} (short-distance) terms.

\begin{figure}
\begin{center}
\epsfig{file=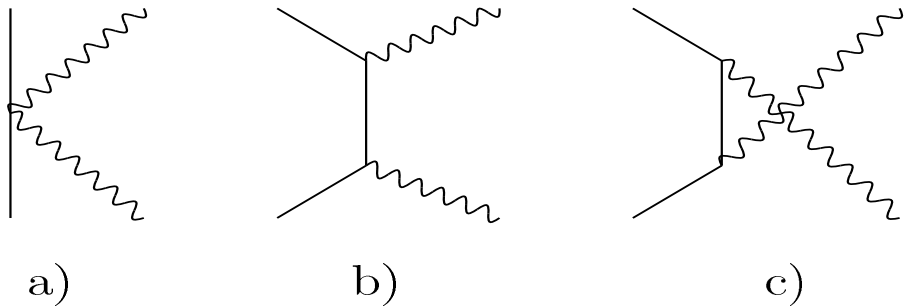,height=4cm,width=12cm}
\end{center}
\caption{{Shown are the a) seagull, b) Born, and c) Cross-Born diagrams contributing to Compton scattering. Here the solid lines represent massive
scalars while the wiggly lines are photons.}}\label{fig:compton}
\end{figure}

One could perform the integration in Eq. \ref{eq:xv} to yield an amplitude which is guaranteed to possess the proper nonanalytic structure.  However, an alternative procedure was posited by Feinberg and Sucher in \cite{Fei88} and involves use of the discontinuity equation, Eq. \ref{eq:xz}, as an integrand in a dispersive integration over the two-photon t-channel cut and a related procedure was recently adapted to the case of gravitational scattering by Bjerrum-Bohr et al~\cite{Bje14}.  In order to carry out this calculation, an analytic continuation is required, since the cross channel annihilation or production amplitude utilized in the dispersive procedure is below threshold for the Compton annihilation and production amplitudes---$t<4m_A^2,\,4m_B^2$.

\begin{figure}
\begin{center}
\epsfig{file=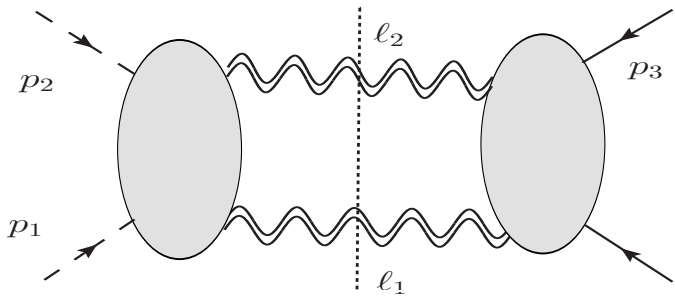,height=4cm,width=10cm}
\caption{\small The two photon cut for the amplitude between a massive
scalar particles. The  grey blobs represent tree-level Compton amplitudes.}
\end{center}
\end{figure}

It is convenient to use the helicity formalism~\cite{Jac59}, where helicity is defined as the projection of the photon spin on the momentum axis.  The helicity amplitudes for the t-channel spin-0-spin-0 Compton annihilation are found in the center of mass frame to have the form~\cite{Hol06}
\begin{eqnarray}
^AA_0^{EM}(++)&=&^AA_0^{EM}(--)=2e^2\left({m_A^2\over E_A^2-\boldsymbol{p}_A^2\cos^2\theta_A}\right)\,,\nonumber\\
^AA_0^{EM}(+-)&=&^AA_0^{EM}(-+)=2e^2\left({\boldsymbol{p}_A^2\sin^2\theta_A\over E_A^2-\boldsymbol{p}_A^2\cos^2\theta_A}\right)\,,
\end{eqnarray}
where $E_A,\, \pm\boldsymbol{p}_A$ are the energy, momentum of the spinless particles and $\theta_A$ is the scattering angle, {\it i.e.} the angle of the outgoing photon $\hat{\boldsymbol{\ell}}_1$ with respect to $\hat{\boldsymbol{p}}_A$---$\cos\theta_A=\hat{\boldsymbol{p}}_A\cdot\hat{\boldsymbol{\ell}}_1$.   It was shown by Feinberg and Sucher that the annihilation amplitudes $A+A'\rightarrow\gamma_1+\gamma_2$ needed in the unitarity relation---Eq. \ref{eq:mk}---can be generated via analytic continuation to imaginary momentum $\boldsymbol{p}_A\rightarrow im_A\xi_A\hat{\boldsymbol{p}}_A$, where $\xi_A^2=1-{t\over 4m_A^2}$ and $t=(p_A-p_A')^2=-4\boldsymbol{p}_A^2$ is the momentum transfer~\cite{Fei88}.  Then
\begin{eqnarray}
^AA_0^{EM}(++)&=&^AA_0^{EM}(--)= 2e^2{1+\tau_A^2\over d_A}\,,\nonumber\\
^AA_0^{EM}(+-)&=&^AA_0^{EM}(-+)= 2e^2{1-x_A^2\over d_A}\,,\label{eq:bv}
\end{eqnarray}
where $\tau_A=\sqrt{t}/2m_A\xi_A$, $x_A=\hat{\boldsymbol{p}}_A\cdot\hat{\boldsymbol{\ell}}_1$, and $d_A=\tau_A^2+x_A^2$.
Equivalently Eq. \ref{eq:bv} can be represented succinctly by
\begin{equation}
^AA_0^{EM}(ab)=2e^2{\cal O_A}^{ij}{\epsilon_{1i}^{a*}}{\epsilon_{2j}^{b*}}\end{equation}
with
\begin{equation}
{\cal O}_A^{ij}={1\over d_A}\left(d_A\delta^{ij}+2\hat{p}_A^i\hat{p}_A^j\right)
\end{equation}
Similarly in the outgoing channel $\gamma_1+\gamma_2\rightarrow B+B'$, again for a spinless particle of charge $e$ but now with mass $m_B$, we define the corresponding quantities $\tau_B=\sqrt{t}/2m_B\xi_B$, $x_B=\hat{\boldsymbol{p}}_B\cdot\hat{\boldsymbol{\ell}}_1$, and $d_B=\tau_B^2+x_B^2$ and the helicity amplitudes can be described by
\begin{equation}
(^BA^{em}_0(cd))^*=2e^2\epsilon_{1k}^{c}\epsilon_{2\ell}^{d}{\cal O_B}^{*k\ell}
\end{equation}
where
\begin{equation}
{\cal O}_B^{k\ell}={1\over d_B}\left(d_B\delta^{k\ell}+2\hat{p}_B^k\hat{p}_B^\ell\right)
\end{equation}
Substituting in Eq. (\ref{eq:xz}), we determine the discontinuity for scattering of spinless particles having masses $m_A,m_B$ across the t-channel two-photon cut for scattering in the CM frame, {\it cf.} Figure 2,
\begin{eqnarray}
{\rm Disc}\,{\rm Amp}_2^{em}(q)&=&-{i\over 2!}{(2e^2)^2\over 4m_Am_B}\int {d^3k_1\over (2\pi)^32k_{10}}{d^3k_2\over (2\pi)^32k_{20}}(2\pi)^4\delta^4(p_1+p_2-k_1-k_2)\nonumber\\
&\times&\sum_{a,b=1}^2\left[{\cal O}_A^{ij}\epsilon^{a*}_{1i}\epsilon^{b*}_{2j}\epsilon^a_{1k}\epsilon^b_{2\ell}{\cal O}_B^{k\ell*}\right]\nonumber\\
&=&-i{e^4\over 16\pi m_Am_B}<\sum_{i,j,k,\ell=1}^3{\cal O}_A^{ij}\delta^T_{ik}\delta^T_{j\ell}{\cal O}_B^{k\ell*}>\nonumber\\
\end{eqnarray}
where
\begin{equation}
\delta^T_{ik}=\sum_{a=1}^2\epsilon_i^{a*}\epsilon_k^{a}=\delta_{ik}-\hat{k}_i\hat{k}_k
\end{equation}
represents the sum over photon polarizations and
$$<G>\equiv \int{d\Omega_{\hat\,{\boldsymbol{k}}}\over 4\pi}G$$
defines the solid-angle average.  Performing the polarization sums, we find
\begin{eqnarray}
&&<\sum_{i,j,k,\ell=1}^3{\cal O}_A^{ij}\delta^T_{ik}\delta^T_{j\ell}{\cal O}_B^{k\ell*}>=<{1\over d_Ad_B}\left[4(y-x_Ax_B)^2-2(1-x_A^2)(1-x_B^2)\right.\nonumber\\
&+&\left.2(1+\tau_A^2)(1+\tau_B^2)\right]>=(4y^2+2\tau_A^2+2\tau_B^2+2\tau_A^2\tau_B^2)I_{00}-8yI_{11}\nonumber\\
&+&2I_{20}+2I_{02}+2I_{22}\label{eq:gb}
\end{eqnarray}
where
$$I_{mn}=<{x_A^mx_B^n\over d_Ad_B}>$$
represents a solid-angle averaged integral and
\begin{equation}
y(s,t)=\hat{\boldsymbol{p}}_A\cdot\hat{\boldsymbol{p}}_B={2s+t-2m_A^2-2m_B^2\over 4m_A\xi_Am_B\xi_B}.
\end{equation}
characterizes the angle between incoming and outgoing spinless particles.  The angle-averaged quantities $I_{00},\,I_{11}$ are have been evaluated by Feinberg and Sucher~\cite{Fei70}. The remaining $I_{mn}$ can be simplified, by repeated use of the algebraic identity $x_i^2=d_i-\tau_i^2$, into elementary forms written in terms of only the fundamental "seed" integrals $I_{00},\,I_{11}$ together with
$$J^A_{00}=<{1\over d_A}>,\,J^B_{00}=<{1\over d_B}>,\,\,{\rm and} <1>$$
and the results are given in Appendix A.  Using this decomposition, we determine then
\begin{eqnarray}
{\rm Disc}\,{\rm Amp}_2^{em}(q)&=&-i{e^4\over 8\pi m_Am_B}\left[2(y^2+\tau_A^2\tau_B^2)I_{00}-4yI_{11}+(1-\tau_A^2)J^A_{00}\right.\nonumber\\
&+&\left.(1-\tau_B^2)J^B_{00}+1\right]\label{eq:vc}
\end{eqnarray}
as the general form for spinless two-particle elastic electromagnetic scattering.  Having obtained this formal result, we can apply Eq. (\ref{eq:vc}) to situations of interest.

\subsection{Massive Particle Electromagnetic Scattering}

We begin with the case of the scattering of two massive spinless particles $A$ and $B$, considered in~\cite{Fei88}. Since we need consider only the small-$t$ component of the scattering amplitude in order to generate the leading large-r behavior, we can use the simplifications $\tau_A,\,\tau_B<<1$ to write
\begin{equation}
{\rm Disc}\,{\rm Amp}_2^{em}(q)\stackrel{t<<m_A^2,m_B^2}{\longrightarrow}-i{e^4\over 8\pi m_Am_B}\left[2y^2I_{00}-4yI_{11}+J^A_{00}+J^B_{00}+1\right]\label{eq:rc}
\end{equation}
There exist two ways to proceed and we consider both techniques, since they can each be useful in specific situations:
\begin{itemize}
\item[a)]  {\bf Direct Evaluation:}  Feinberg and Sucher evaluated the angular integrals directly to yield the scattering amplitude~\cite{Fei88}. That is, when $t<<m_A^2,\,m_B^2$ we have, as shown in Appendix C
\begin{equation}
\left(
\begin{array}{c}
I_{00}\\
I_{11}\\
J^A_{00}\\
J^B_{00}
\end{array}
\right)\stackrel{t<<m_A^2,\,m_B^2}{\longrightarrow}
\left(
\begin{array}{c}
-{1\over 3}+i2\pi{m_rm_Am_B\over p_0t}\\
-1+i\pi{m_r\over 2p}\\
{\pi m_A\over \sqrt{t}}-1\\
{\pi m_B\over \sqrt{t}}-1\\
\end{array}\right)+{\cal O}(\sqrt{t})
\end{equation}
where $p_0=\sqrt{{m_r(s-s_0)\over m_A+m_B}}$ is the center of mass momentum for the spinless scattering process with $m_r=m_Am_B/(m_A+m_B)$ being the reduced mass.
Also, in this limit we have, near scattering threshold---$s\simeq s_0=(m_A+m_B)^2$---
\begin{equation}
y(s_0,t)={2s_0+t-2m_A^2-2m_B^2\over 4m_A\xi_Am_B\xi_B}=1+\ldots
\end{equation}
whereby Eq. (\ref{eq:rc}) becomes
\begin{eqnarray}
&&{\rm Disc}\,{\rm Amp}_2^{em}(q)\stackrel{t<<m_A^2,\,m_B^2}{\longrightarrow}-i{e^4\over 8\pi m_Am_B}\left[-{2\over 3}+4-1-1+1+{\pi(m_A+m_B)\over \sqrt{t}}\right.\nonumber\\
&+&\left.i4\pi{m_rm_Am_B\over p_0t}\right]=-i{e^2\over 8\pi m_Am_B}\left({7\over 3}+{\pi(m_A+m_B)\over \sqrt{t}}+i4\pi{m_rm_Am_B\over p_0t}\right)\nonumber\\
\quad
\end{eqnarray}
Defining $L=\log(-t)$ and $S=\pi^2/\sqrt{-t}$ and noting that
\begin{equation}
{\rm Disc}\,\left[\log(-t),\,\sqrt{1\over -t}\right]=\left[2\pi i,\,-i{2\pi^2\over \sqrt{t}}\right]\,,
\end{equation}
the scattering amplitude is given by
\begin{equation}
{\rm Amp}_2^{em}(q)=-{e^4\over 16\pi^2m_Am_B}\left[{7\over 3}L-S(m_A+m_B)+4\pi i{m_rm_Am_B\over p_0t}L+\ldots\right]\,,\label{eq:wc}
\end{equation}

The imaginary component of Eq. (\ref{eq:wc}) represents the Coulomb phase, or equivalently the contribution of the second Born approximation, which must be subtracted in order to define a proper higher-order potential.  Using~\cite{Dal51}
\begin{eqnarray}
B_2^{em}(q)&=&i\int{d^3\ell\over (2\pi)^3}{e^2\over |\boldsymbol{p}_f-\boldsymbol{\ell}|^2+\lambda^2}
{i\over {p_0^2\over 2m_r}-{\ell^2\over 2m_r}+i\epsilon}{e^2\over |\boldsymbol{\ell}-\boldsymbol{p}_i|^2+\lambda^2}\nonumber\\
&=&-i{e^4\over 4\pi}{m_r\over p_0}{L\over t}\,,\label{eq:cx}
\end{eqnarray}
and subtracting, what remains is the higher order electromagnetic amplitude we are seeking.  Then, writing $t=q^2$ and taking the nonrelativistic limit, the Fourier transform yields the second-order effective electromagnetic potential
\begin{eqnarray}
V_2^{em}(r)&=&-\int{d^3q\over (2\pi)^3}e^{-i\boldsymbol{q}\cdot\boldsymbol{r}}\left({\rm Amp}_2^{em}(q)-B_2^{em}(q)\right)\,,\nonumber\\
&=&-{\alpha_{em}^2(m_A+m_B)\over 2 m_Am_Br^2}-{7\alpha_{em}^2\hbar\over 6\pi m_Am_B r^3}\label{eq:dz}
\end{eqnarray}

In comparing with previous calculations, it is necessary to understand an important point made by Sucher~\cite{Suc94}, which is that the result for the {\it classical} component of Eq. (\ref{eq:dz}) depends on the specific form of the lowest order potential and the propagator used to generate the Born subtraction. The result $V_2^{em}(r)$ follows from use of the simplest nonrelativistic forms for each. Inclusion of relativistic corrections in either the potential or the propagator (or both) will yield the same imaginary piece as found in Eq. (\ref{eq:cx}) but also in general a term involving $\pi^2\over \sqrt{-t}$, generating a correction to the classical potential, while the quantum piece is unchanged.  Thus the effective potential quoted in Eq. (\ref{eq:dz}) agrees with the previous result found by Ross and Holstein~\cite{Ros08} but {\it not} with the results of Feinberg and Sucher~\cite{Fei88}, of Iwasaki~\cite{Iwa71}, or of Spruch~\cite{Spr93}.  What is {\it identical} in each calculation is the form of the on-shell scattering amplitude
\begin{equation}
{\rm Amp}_2^{em}(q)=\int d^3r e^{-i\boldsymbol{q}\cdot\boldsymbol{r}}(V_0^{em}(r)+V_2^{em}(r))+B_2^{em}(q)
\end{equation}

\item[b)] {\bf Feynman Integral Technique:}  It is useful to take an alternate tack, as pursued in \cite{Bje14}, wherein one writes the quantities $I_{mn}$ in terms of the discontinuity of familiar Feynman scalar integrals over the two-photon or two-graviton t-channel cut.  As shown in Appendix B, these relations are
\begin{eqnarray}
I_{00}&=&<{1\over d_Ad_B}>=-16\pi m_A^2\xi_A^2m_B^2\xi_B^2\,{\rm Disc}\,(I_4(s,t)+I_4(u,t))\nonumber\\
I_{11}&=&\,<{x_Ax_B\over d_Ad_B}>=4\pi tm_A\xi_Am_B\xi_B\,{\rm Disc}\,(I_4(s,t)-I_4(u,t))\nonumber\\
J^A_{00}&=&<{1\over d_A}>=-8\pi m_A^2\xi_A^2\,{\rm Disc}\,(I_3(p_1,q,m_A)+I_3(p_2,q,m_A))\nonumber\\
J^B_{00}&=&<{1\over d_A}>=-8\pi m_B^2\xi_B^2\,{\rm Disc}\,(I_3(-p_3,q,m_A)+I_3(-p_4,q,m_A))\nonumber\\
1&=&<1>=\,8\pi\,{\rm Disc}\,I_2(q)
\end{eqnarray}
where
\begin{eqnarray}
I_4(s,t)&\equiv&{\rm Amp}_{box}=\int{d^4\ell\over (2\pi)^4}{1\over \ell^2(\ell-q)^2}\left({1\over ((\ell-p_1)^2-m_A^2)((\ell-p_3)^2-m_B^2)}\right)\nonumber\\
&=&\int{d^4\ell\over (2\pi)^4}{1\over \ell^2(\ell-q)^2}\left({1\over ((\ell-p_2)^2-m_A^2)((\ell-p_4)^2-m_B^2)}\right)\nonumber\\
I_4(u,t)&\equiv&{\rm Amp}_{c-box}=\int{d^4\ell\over (2\pi)^4}{1\over \ell^2(\ell-q)^2}\left({1\over ((\ell-p_1)^2-m_A^2)((\ell-p_4)^2-m_B^2)}\right)\nonumber\\
&=&\int{d^4\ell\over (2\pi)^4}{1\over \ell^2(\ell-q)^2}\left({1\over ((\ell-p_2)^2-m_A^2)((\ell-p_3)^2-m_B^2)}\right)
\end{eqnarray}
are scalar box and cross-box integrals---{\it cf.} Figure 3d,\,3e,
\begin{eqnarray}
I_3(p_1,q,m_A)&=&{\rm Amp}_{triangle}=\int{d^4\ell\over (2\pi)^4}{1\over \ell^2(\ell-q)^2((\ell-p_1)^2-m_A^2)}\nonumber\\
I_3(-p_3,q,m_B)&=&{\rm Amp}_{triangle}=\int{d^4\ell\over (2\pi)^4}{1\over \ell^2(\ell-q)^2((\ell+p_3)^2-m_B^2)}\nonumber\\
\quad
\end{eqnarray}
are scalar triangle diagrams---{\it cf.} Figure 3b,\,3c, and
\begin{equation}
I_2(q)={\rm Amp}_{bubble}=\int{d^4\ell\over (2\pi)^4}{1\over \ell^2(\ell-q)^2}
\end{equation}
is the scalar bubble---{\it cf.} Figure 3a.

 \begin{figure}
\begin{center}
\epsfig{file=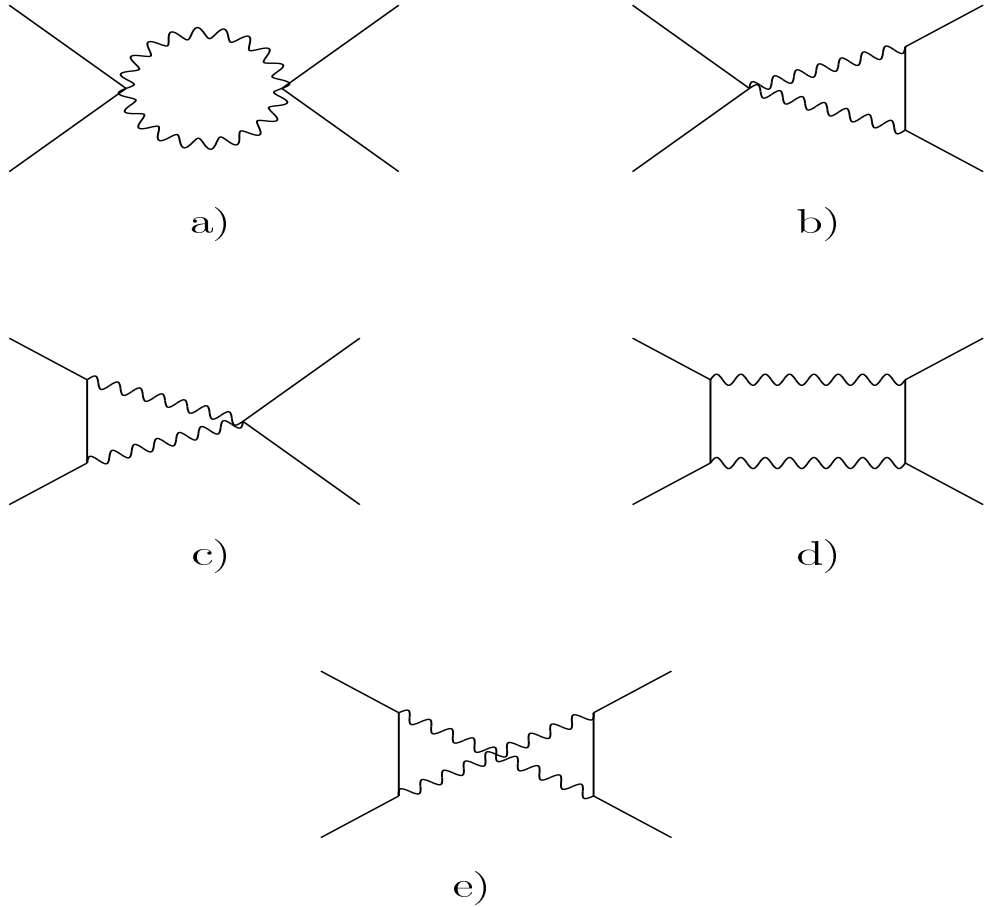,height=12cm,width=13cm}
\caption{{Shown are the a) bubble, b),c) triangle, d) Box and e) Cross-box diagrams contributing to spinless particle scattering.  Here the solid lines designate the massive spinless particles, while the wiggly lines represent photons in the case of the electromagnetic interaction or gravitons in the case of gravitational scattering.}}
\end{center}
\end{figure}

As shown in Appendix A, simple algebraic identities can be used to write any of the angular integrals $I_{nm}$ in terms of linear combinations of these five basic Feynman integrals.  In this way the scattering amplitude discontinuity given in Eq. (\ref{eq:vc}) becomes
\begin{eqnarray}
&&{\rm Disc}\,{\rm Amp}_2^{em}(q)=i{e^4\over 8\pi m-Am_B}\left[2y^2\cdot 16\pi m_A^2m_B^2(I_4(s,t)+I_4(u,t))\right.\nonumber\\
&+&\left.4y\cdot4\pi m_Am_Bt(I_4(s,t)-I_4(u,t))+8\pi m_A^2(I_3(p_1,q,m_A)+I_3(p_2,q,m_A))\right.\nonumber\\
&+&\left.8\pi m_B^2(I_3(-p_3,q,m_B)+I_3(-p_4,q,m_B))-8\pi I_2(q)\right]
\end{eqnarray}
The Feynman integrals are well known~\cite{Don94}
\begin{equation}
\left(\begin{array}{c}
I_4(s,t)\\
I_4(u,t)\\
I_3(p_1,q,m_A)\\
I_3(-p_3,q,m_B)\\
I_2(q)
\end{array}\right)
\stackrel{t<<m_A^2,\,m_B^2}{\longrightarrow}
{i\over 32\pi^2}\left(\begin{array}{c}
{2\over m_Am_Bt}(1+{s-s_0\over 6m_Am_B})L+i4\pi{L\over m_A^2m_B^2t}\sqrt{m_Am_B\over s-s_0}\\
-{2\over m_Am_Bt}(1+{s+t-s_0\over 6m_Am_B})L\\
-{1\over m_A^2}(L+Sm_A)\\
-{1\over m_B^2}(L+Sm_B)\\
-2L
\end{array}\right)+\ldots
\end{equation}
and, using the near-threshold identity
\begin{equation}
y(s_0,t)\stackrel{t<<m_A^2,\,m_B^2}{\longrightarrow}1+\ldots
\end{equation}
we find directly
\begin{eqnarray}
&&{\rm Amp}_2^{em}(q)\stackrel{t<<m_A^2,m_B^2}{\longrightarrow}{ie^4\over 4m_Am_B}\left[16m_A^2m_B^2(I_4(s,t))+I_4(u,t))\right.\nonumber\\
&+
&\left.8m_Am_Bt(I_4(s,t)-I_4(u,t))\right.\nonumber\\
&+&\left.4m_A^2(I_3(p_1,q,m_A)+I_3(p_2,q,m_A))\right.\nonumber\\
&+&\left.4m_B^2(I_3(-p_3,q,m_A)+I_3(-p_4,q,m_A))-4I_2(q)\right]\nonumber\\
&=&-{e^4\over 128\pi^2m_Am_B}\left[16\cdot\left(-{1\over 3}L+i4\pi {L\over t}\sqrt{m_Am_B\over s-s_0}\right)+8\left(4L+i4\pi{L\over m_Am_B}\sqrt{m_Am_B\over s-s_0}\right)\right.\nonumber\\
&+&\left.8\left(-Sm_A-L\right)+8\left(-Sm_B-L\right)-4\cdot-2L\right]\nonumber\\
&=&-{e^4\over 16\pi^2m_Am_B}\left[{7\over 3}L-S(m_A+m_B)+4\pi i{m_rm_Am_B\over p_0t}L+\ldots\right]
\end{eqnarray}
which is identical to the result obtained in Eq. (\ref{eq:wc}).
\end{itemize}

\subsection{Massive-Massless Electromagnetic Scattering}

Recently, as a model for quantum-mechanical light-bending in the vicinity of a massive object, the scattering of massless and massive spin-zero systems was considered~\cite{Bje15}, so it is interesting to first treat the simpler case of the electromagnetic scattering of a massless spinless particle $A$ by a massive spinless particle $B$. Again, we can use either procedure described above, but certain changes are required.
\begin{itemize}
\item[a)] {\bf Direct Evaluation:}   Since Eq. (\ref{eq:vc}) is exact, this form can still be used, but kinematic modifications which arise in the $m_A\rightarrow 0$ limit must be invoked. Writing $s-m_B^2=2m_BE$, where $E$ is the laboratory frame energy of the incident massless particle, we now divide by the normalizing factor $4Em_B$. Also, since $m_A\xi_A\rightarrow -i\sqrt{t}/2$ we find $\tau_A\rightarrow i$ and so
    \begin{eqnarray}
{\rm Disc}\,{\rm Amp}_2^{0-em}(q)&=&-i{e^4\over 8\pi Em_B}\left[2(y^2-\tau_B^2)I^{(0)}_{00}-4yI^{(0)}_{11}+2J^{A{(0)}}_{00}\right.\nonumber\\
&+&\left.(1-\tau_B^2)J^B_{00}+1\right]\label{eq:vd}
\end{eqnarray}
where we have included a superscript $(0)$ to indicate the modified forms required when $m_A\rightarrow 0$.  Another important change is the form of the quantity $y$, which becomes imaginary
\begin{equation}
y\stackrel{m_A\rightarrow 0}{\longrightarrow} i{s-m_B^2+{t\over 2}\over \sqrt{t}m_B\xi_B}=i|y|\quad{\rm with}\quad |y|\simeq i{2E\over \sqrt{t}}
\end{equation}
We make the small angle scattering approximation---$\sqrt{t}<<2E$---so that $|y|>>1$.   Since also $t<<m_B^2$, we find the simplified form
\begin{eqnarray}
&&{\rm Disc}\,{\rm Amp}_2^{0-em}(q)\stackrel{|y|>>1}{\longrightarrow}-i{e^4\over 8\pi Em_B}\left[2y^2I^{(0)}_{00}-4yI^{(0)}_{11}\right.\nonumber\\
&+&\left.2J^{A{(0)}}_{00}+J^B_{00}+1\right]\label{eq:nq}
\end{eqnarray}
The massless seed integrals are evaluated in Appendix C, yielding
\begin{eqnarray}
I^{(0)}_{00}&=&m_B^2\xi_B^2\left[{1\over s-m_B^2}\ln{s-m_B^2\over m_B^2\xi_B^2}+{1\over u-m_B^2}\ln{u-m_B^2\over m_B^2\xi_B^2}+\ldots\right]\nonumber\\
&\simeq& {m_B\over 2E}\ln \left({2E\over m_B}\right)-{m_B\over 2E+{t\over m_B}}\ln -\left({2E\over m_B}+{t\over m_B^2}\right)\nonumber\\
I^{(0)}_{11}&=&{i\sqrt{t}\over 2}m_B\xi_B\left[{1\over s-m_B^2}\ln{s-m_B^2\over m_B^2\xi_B^2}-{1\over u-m_B^2}\ln{u-m_B^2\over m_B^2\xi_B^2}+\ldots\right]\nonumber\\
&\simeq&i{\sqrt{t}\over 2}\left[{1\over 2E}\ln\left({2E\over m_B}\right)+{1\over 2E+{t\over m_B^2}}\ln -\left({2E\over m_B}+{t\over m_B^2}\right)\right] \nonumber\\
J^{A(0)}_{00}&=&-{1\over 2}\ln{-t\over \lambda^2}
\end{eqnarray}
where $\lambda$ is a cutoff introduced to regularize the triangle integral and we have omitted ultraviolet divergent pieces, which are absorbed into renormalized coefficients of short distance terms.  We have then
\begin{eqnarray}
{\rm Disc}\,{\rm Amp}_2^{0-em}(q)&\simeq& -i{e^4\over 8\pi Em_B}\left[-{2(s+{t\over 2}-m_B^2)^2\over t}\left[{1\over s-m_B^2}\ln{s-m_B^2\over m_B^2}\right.\right.\nonumber\\
&+&\left.\left.{1\over u-m_B^2}\ln{u-m_B^2\over m_B^2}\right]\right.\nonumber\\
&+&\left.2(s+{t\over 2}-m_B^2)\left[{1\over s-m_B^2}\ln{s-m_B^2\over m_B^2}-{1\over u-m_B^2}\ln{u-m_B^2\over m_B^2}\right]\right.\nonumber\\
&-&\left.\ln{-t\over \lambda^2}+{\pi m_B\over \sqrt{t}}\right]
\end{eqnarray}
Using
\begin{equation}
(s-m_B^2)^2,\,(u-m_B^2)^2=(s+{t\over 2}-m_B^2\mp{t\over 2})^2\simeq (s+{t\over 2}-m_B^2)^2\mp t(s+{t\over 2}-m_B^2)
\end{equation}
we find
\begin{eqnarray}
&&{\rm Amp}_2^{0-em}(q)={e^4\over 16\pi^2 Em_B}\left\{{2L\over t}\left[(s-m_B^2)\ln{s-m_B^2\over m_B^2}+(u-m_B^2)\ln{u-m_B^2\over m_B^2}\right]\right.\nonumber\\
&+&\left.{1\over 2}L^2+Sm_B\right\}\nonumber\\
&=&{e^4\over 16\pi^2 Em_B}\left\{{2L\over t}\left[2m_BE\ln\left({2E\over m_B}\right)-(2E+{t\over m_B})\ln-\left({2E\over m_B}+{t\over m_B^2}\right)\right]\right.\nonumber\\
&+&\left.{1\over 2}L^2+Sm_B\right\}
\label{eq:bz}
\quad
\end{eqnarray}
Here a double logarithm has appeared, as is common in the presence of a vanishing mass.

\item[b)] {\bf Feynman Integral Technique:} Equivalently, we can express Eq. (\ref{eq:nq}) as
\begin{eqnarray}
&&{\rm Disc}\,{\rm Amp}_2^{0-em}(q)\stackrel{m_A\rightarrow 0}{\longrightarrow}-i{e^4\over 8\pi Em_B}{\rm Disc}\,\left[-8\pi(s+{t\over 2}-m_B^2)^2(I^{(0)}_4(s,t)+I_4^{(0)}(u,t))\right.\nonumber\\
&+&\left.8\pi t(s+{t\over 2}-m_B^2)(I^{(0)}_4(s,t)-I_4^{(0)}(u,t))+4\pi t(I_3^{(0)}(p_1,q,m_A)+I_3^{(0)}(p_2,q,m_A))\right.\nonumber\\
&-&\left.8\pi m_B^2(I_3(-p_3,q,m_B)+I_3(-p_4,q,m_B))+8\pi I_2(q)\right]
\end{eqnarray}
Using, from Appendix C,
\begin{eqnarray}
I_4^{(0)}(s,t)+I_4^{(0)}(u,t)&=&{iL\over 8\pi^2t}\left[{1\over s-m_B^2}\ln{s-m_B^2\over m_B^2}+{1\over u-m_B^2}\ln{u-m_B^2\over m_B^2}+\ldots\right]\nonumber\\
I_4^{(0)}(s,t)-I_4^{(0)}(u,t)&=&{iL\over 8\pi^2t}\left[{1\over s-m_B^2}\ln{s-m_B^2\over m_B^2}-{1\over u-m_B^2}\ln{u-m_B^2\over m_B^2}+\ldots\right]\nonumber\\
I_3^{(0)}(p_1,q,m_A)&=&{i\over 32\pi^2t}L\,,
\end{eqnarray}
we have
\begin{eqnarray}
&&{\rm Amp}_2^{0-em}(q)={e^4\over 16\pi^2Em_B}\left\{{2L\over t}\left[(s-m_B^2)\ln{s-m_B^2\over m_B^2}+(u-m_B^2)\ln{u-m_B^2\over m_B^2}\right]\right.\nonumber\\
&+&\left.{1\over 2}L^2+Sm_B\right\}\nonumber\\
&=&{e^4\over 16\pi^2Em_B}\left\{{2L\over t}\left[2m_BE\ln\left({2E\over m_B}\right)-(2E+{t\over m_B})\ln-\left({2E\over m_B}+{t\over m_B^2}\right)\right]\right.\nonumber\\
&+&\left.{1\over 2}L^2+Sm_B\right\}
\label{eq:bk}
\quad
\end{eqnarray}
as before, {\it cf.} Eq. (\ref{eq:bz}).
\end{itemize}

Defining
\begin{eqnarray}
{\rm Amp}_2^{0-em}(q)&\equiv&{e^4\over 4Em_B}\left[ao^\phi(s,t)I_4^{(0)}(s,t)+ao^\phi(u,t)I_4^{(0)}(u,t)+{1\over 2}s_{12}^\phi(I_3^{(0)}(p_1,q,0)\right.\nonumber\\
&+&\left.I_3^{(0)}(p_2,q,0))+{1\over 2}s_{34}^\phi(I_3(-p_3,q,m_B)+I_3(-p_4,q,m_B))+au^\phi I_2(q)\right]\nonumber\\
\quad
\end{eqnarray}
and using the result that $u-m_B^2\stackrel{t\rightarrow 0}{\longrightarrow} -(s-m_B^2)$, we find
\begin{equation}
{ao^\phi(s,t)\over s-m_B^2}+{ao^\phi(u,t)\over u-m_B^2}=-4i\left[(s-m_B^2)+(u-m_B^2)\right]\stackrel{t\rightarrow 0}{\longrightarrow}4it=-s_{12}^\phi
\end{equation}
so that the BCJ relation is satisfied and serves as a useful check on our result~\cite{Ber08}.  We also note that
\begin{eqnarray}
&&\left[(s-m_B^2)\ln{s-m_B^2\over m_B^2}+(u-m_B^2)\ln{u-m_B^2\over m_B^2}\right]\nonumber\\
&&\stackrel{t\rightarrow 0}{\longrightarrow}(s-m_B^2)\left[\ln{s-m_B^2\over m_B^2}-\ln -\,{s-m_B^2\over m_B^2}\right]\nonumber\\
&=&-i\pi(s-m_B^2)
\end{eqnarray}
so that the scattering amplitude picks up an imaginary component which can be identified as a Coulomb scattering phase as before and thereby subtracted off. In terms of the laboratory energy $E$, we have then the effective potential
\begin{eqnarray}
V_2^{0-em}(r)&=&-\int{d^3q\over (2\pi)^3}e^{-i\boldsymbol{q}\cdot\boldsymbol{r}}\left({\rm Amp}_2^{0-em}(q)-B_2^{0-em}(q)\right)\,,\nonumber\\
&=&{\alpha_{em}^2\over 2Er^2}+{2\alpha_{em}^2\hbar\over \pi m_BEr^3}\ln{r\over r_0}
\end{eqnarray}

We see then that the case of electromagnetic scattering of spinless particles can be straightforwardly and simply treated via on-shell methods, and move to our primary goal, which is gravitational scattering.

\section{On-Shell Method: Gravitational Scattering}

We now consider the analogous gravitational calculation.  In this case the Feynman diagram calculation is considerably more challenging than its electromagnetic analog.  The reasons for this are at least two.  One is the replacement of the electromagnetic interaction by its gravitational analog
\begin{equation}
{\cal L}_{int}={\kappa\over 2}h_{\mu\nu} T^{\mu\nu}
\end{equation}
where $\kappa=\sqrt{32\pi G}$ is the gravitational coupling, $g_{\mu\nu}\equiv\eta_{\mu\nu}+h_{\mu\nu}$ is the gravitational metric tensor, and $T^{\mu\nu}$ is the energy-momentum tensor, which at leading order has the matrix element between spinless particles
\begin{equation}
<p_2|T^{\mu\nu}(x)|p_1>=2(p_1+p_2)_\mu(p_1+p_2)_\nu e^{i(p_2-p_1)\cdot x}+{\cal O}(q)
\end{equation}
Thus one deals with the energy-momentum tensor rather than the familiar electromagnetic current.  The second increase in complexity can be seen from the form of the gravitational Compton scattering amplitude needed for the on=shell technique
\begin{eqnarray}
{\rm Amp}_{GC}(S=0)&=&\kappa^2 \left\{-{(\epsilon_i\cdot
p_i)^2(\epsilon_f^*\cdot p_f)^2\over p_i\cdot k_i}+{(\epsilon_f^*\cdot
p_i)^2(\epsilon_i\cdot
p_f)^2\over p_i\cdot k_f}\right.\nonumber\\
&+&\left.\left[\epsilon_f^*\cdot\epsilon_i(\epsilon_i\cdot
p_i\epsilon_f^*\cdot p_f+\epsilon_i\cdot p_f\epsilon_f^*\cdot
p_i)-{1\over 2}k_i\cdot
k_f(\epsilon_f^*\cdot\epsilon_i)^2\right]\right.\nonumber\\
&+&\left.{1\over 2k_i\cdot k_f}
\left[\epsilon_f^*\cdot p_f\epsilon_f^*\cdot
p_i(\epsilon_i\cdot(p_i-p_f))^2\right.\right.\nonumber\\
&+&\left.\left.\epsilon_i\cdot p_i\epsilon_i\cdot
p_f(\epsilon_f^*\cdot(p_i-p_f))^2\right.\right.\nonumber\\
&+&\left.\left.\epsilon_i\cdot(p_i-p_f)\epsilon_f^*\cdot(p_f-p_i)(\epsilon_f^*\cdot
p_f\epsilon_i\cdot p_i+\epsilon_f^*\cdot p_i\epsilon_i\cdot
p_f)\right.\right.\nonumber\\
&-&\left.\left.\epsilon_f^*\cdot\epsilon_i\left(\epsilon_i\cdot(p_i-p_f)\epsilon_f^*\cdot
(p_f-p_i)(p_i\cdot p_f-m^2)\right.\right.\right.\nonumber\\
&+&\left.\left.\left.k_i\cdot k_f(\epsilon_f^*\cdot p_f\epsilon_i\cdot
p_i+\epsilon_f^*\cdot p_i\epsilon_i\cdot p_f)\right.\right.\right.\nonumber\\
&+&\left.\left.\left.\epsilon_i\cdot(p_i-p_f)(\epsilon_f^*\cdot p_fp_i\cdot
k_f+\epsilon_f^*\cdot p_ip_f\cdot k_f)\right.\right.\right.\nonumber\\
&+&\left.\left.\left.\epsilon_f^*\cdot(p_f-p_i)(\epsilon_i\cdot
p_ip_f\cdot
k_i+\epsilon_i\cdot p_fp_i\cdot k_i)\right)\right.\right.\nonumber\\
&+&\left.\left.(\epsilon_f^*\cdot\epsilon_i)^2\left(p_i\cdot k_ip_f\cdot
k_i+p_i\cdot k_fp_f\cdot k_f\right.\right.\right.\nonumber\\
&-&\left.\left.\left.{1\over 2}(p_i\cdot k_ip_f\cdot
k_f+p_i\cdot k_fp_f\cdot k_i)\right.\right.\right.\nonumber\\
&+&\left.\left.\left.{3\over 2}k_i\cdot k_f(p_i\cdot
p_f-m^2)\right)\right]\right\}\label{eq:gx}
\end{eqnarray}
which results from the four diagrams shown in Figure 4.  The additional (graviton-pole) diagram compared to the electromagnetic case involves the triple graviton vertex, which is required due to the nonlinearity of the gravitational interaction.

 \begin{figure}
\begin{center}
\epsfig{file=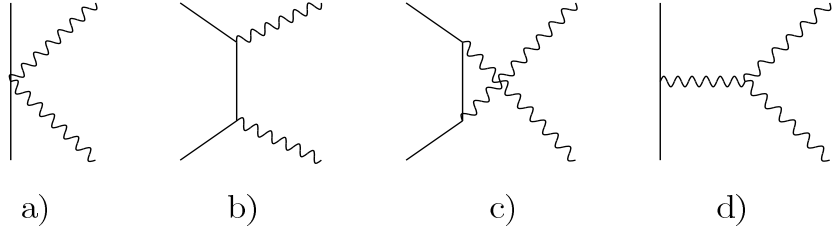,height=3cm,width=11cm}
\caption{{Shown are the a) seagull, b) Born, c) Cross-Born, and d) graviton pole diagrams contributing the gravitational Compton scattering.  Here the olid lines represent massive spinless particles while the wiggly lines are gravitons.}}
\label{fig:gravcomp}
\end{center}
\end{figure}

However, Eq. (\ref{eq:gx}) can be greatly simplified, since it has been pointed out that such gravitational amplitudes
factorize into products of electromagnetic amplitudes times a simple kinematic
factor~\cite{Cho95},\cite{Ber02}
\begin{equation}
F={p_i\cdot k_ip_i\cdot k_f\over k_i\cdot k_f}\,,
\end{equation}
which, in the center of mass frame can be written as
\begin{equation}
F={E^2(E-|\boldsymbol{p}|\cos\theta)(E+|\boldsymbol{p}|\cos\theta)\over 2E^2}={1\over 2}(E^2-\boldsymbol{p}^2\cos^2\theta)
\end{equation}
That is, we have the remarkable identity
\begin{eqnarray}
{\rm Amp}_{C}^{grav}(S=0)&=&{\kappa^2\over 8e^4}F\left({\rm Amp}_C(S=0)\right)^2\nonumber\\
&=&{\kappa^2\over 2}\left({p_i\cdot k_ip_f\cdot k_f\over k_i\cdot k_f}\right)\left({\epsilon_i\cdot p_i\epsilon_f^*\cdot p_f\over p_i\cdot k_i}-{\epsilon_i\cdot p_f\epsilon_f^*\cdot p_i\over p_i\cdot k_f}-\epsilon_i\cdot\epsilon_f^*\right)\nonumber\\
&\times&\left({\epsilon_i\cdot p_i\epsilon_f^*\cdot p_f\over p_i\cdot k_i}-{\epsilon_i\cdot p_f\epsilon_f^*\cdot p_i\over p_i\cdot k_f}-\epsilon_i\cdot\epsilon_f^*\right)
\end{eqnarray}
 For application to gravitational scattering, it is again useful to use the helicity formalism.  Using factorization, the CM helicity amplitudes for spin-0 gravitational Compton annihilation of a particle with mass $m_A$ are found to be
\begin{eqnarray}
^AB_0^{grav}(++)&=&^AB_0^{grav}(--)={\kappa^2\over 4}(E^2-\boldsymbol{p}_A^2x^2)[{m_A^2\over E^2-\boldsymbol{p}_A^2x_A^2}]^2={\kappa^2\over 4}{m_A^4\over E^2-\boldsymbol{p}_A^2x_A^2}\,,\nonumber\\
^AB_0^{grav}(+-)&=&^AB_0^{grav}(-+)={\kappa^2\over 4}(E^2-\boldsymbol{p}_A^2x_A^2)[{\boldsymbol{p}_A^2(1-x_A^2)\over E^2-\boldsymbol{p}_A^2x_A^2}]^2={\kappa^2\over 4}{\boldsymbol{p}_A^4(1-x_A^2)^2\over E^2-\boldsymbol{p}_A^2x_A^2} \,.\nonumber\\
\quad
\end{eqnarray}
Making the analytic continuation, as before, we find
\begin{eqnarray}
^AB_0^{grav}(++)&=&^AB_0^{grav}(--)\stackrel{FS}{\longrightarrow}{\kappa^2m_A^2\xi_A^2\over 4}{(1+\tau_A^2)^2\over d_A}\nonumber\\
^AB_0^{grav}(+-)&=&^AB_0^{grav}(-+)\stackrel{FS}{\longrightarrow}{\kappa^2m_A^2\xi_A^2\over 4}{(1-x_A^2)^2\over d_A}\label{eq:cw}
\end{eqnarray}
and similar forms hold for the final state reaction $\gamma_1+\gamma_2\rightarrow P+P'$ wherein two photons annihilate into a pair of spinless particles with mass $m_B$. We can write these results in the succinct forms
\begin{eqnarray}
^AB_0^{grav}(ij)&=&{\kappa^2m_A^2\xi_A^2d_A\over 4}{\cal O}_A^{rs}\epsilon^{*i}_r\epsilon^{*j}_s{\cal O}_A^{uv}\epsilon^{*i}_u\epsilon^{*j}_v\nonumber\\
^BB_0^{grav}(ij)&=&{\kappa^2m_B^2\xi_B^2d_B\over 4}\epsilon^{i}_r\epsilon^{j}_s{\cal O}_B^{rs}\epsilon^{i}_u\epsilon^{j}_v{\cal O}_B^{uv}
\end{eqnarray}
We have then for the gravitational scattering discontinuity
\begin{eqnarray}
&&{\rm Disc}\,{\rm Amp}_2^{grav}(q)=-{i\over 2!}{1\over 4m_Am_B}\int {d^3k_1\over (2\pi)^32k_1^0}{d^3k_2\over (2\pi)^32k_2^0}(2\pi)^4\delta^4(p_1+p_2-k_1-k_2)\nonumber\\
&\times& \sum_{a,b=1}^2{\kappa^4m_A^2\xi_A^2m_B^2\xi_B^2d_Ad_B\over 16}{\cal O}_A^{ij}\epsilon^{a*}_{1i}\epsilon^{b}_{2j}{\cal O}_A^{k\ell}\epsilon^{a}_{1k}\epsilon^{b*}_{2\ell}\epsilon^a_{1a}\epsilon^b_{2b}{\cal O}_B^{ab*}\epsilon^a_{1c}\epsilon^b_{2\ell}{\cal O}_B^{cd*}\nonumber\\
&=&-i{\kappa^4m_A\xi_A^2m_B\xi_B^2\over 1024\pi}<\sum_{i,j,k,\ell=1}^3\sum_{a,b,c,d=1}^3 d_Ad_B{\cal O}_A^{ij}{\cal O}_A^{k\ell}P^G_{ik;ac}P^G_{j\ell;bd}{\cal O}_B^{ab*}{\cal O}_B^{cd*}>\nonumber\\
\quad
\end{eqnarray}
where, as before, we have defined $\tau_i=\sqrt{t}/2m_i$, $x_i=\hat{\boldsymbol{p}}_i\cdot\hat{\boldsymbol{k}}$, $d_i=\tau_i^2+x_i^2$, with $i=A,B$ and
$$<G>\equiv \int{d\Omega_{\hat{\boldsymbol{k}}}\over 4\pi}\,G\,,$$
while the sum over graviton polarizations is given by
\begin{equation}
P^G_{ik;ac}=\sum_{r=1}^2\epsilon^{r*}_i\epsilon^{r*}_i\epsilon^{r}_a\epsilon^{r}_c={1\over 2}\left[\delta^T_{ia}\delta^T_{kc}+\delta^T_{ic}\delta^T_{ka}-\delta^T_{ik}\delta^T_{ac}\right]
\end{equation}
Performing the polarization sum, we find the (exact) result
\begin{eqnarray}
&&<\left[d_Ad_B{\cal O}_A^{ij}{\cal O}_A^{k\ell}P^G_{ik;ac}P^G_{j\ell;bd}{\cal O}_B^{ab*}{\cal O}_B^{cd*}\right]>=
<{1\over d_Ad_B}\left[4\left(2(y-x_Ax_B)^2\right.\right.\nonumber\\
&-&\left.\left.(1-x_A^2)(1-x_B^2)\right)^2-2(1-x_A^2)^2(1-x_B^2)^2+2(1+\tau_A^2)^2(1+\tau_B^2)^2\right]>\nonumber\\
&=&(16y^4-16y^2+4+4\tau_A^2+4\tau_B^2+8\tau_A^2\tau_B^2+2\tau_A^4+2\tau_B^4+4\tau_A^4\tau_B^2+4\tau_A^2\tau_B^4+2\tau_A^4\tau_B^4)I_{00}\nonumber\\
&+&(32y-64y^3)I_{11}+(16y^2-4)I_{20}+(16y^2-4)I_{02}+(80y^2-8)I_{22}+2I_{40}+2I_{04}\nonumber\\
&-&32yI_{31}-32yI_{13}+12I_{42}+12I_{24}-32yI_{33}+2I_{44}\label{eq:nb}
\end{eqnarray}
Using the results of Appendix A for the solid-angle averaged integrals $I_{mn}$, we have then
\begin{eqnarray}
{\rm Disc}\,{\rm Amp}_2^{grav}(q)&=&-i{\kappa^4m_A\xi_A^2m_B\xi_B^2\over 1024\pi}\left[4I_{00}\left(4y^4-4y^2(1+\tau_A^2+\tau_B^2)+1\right.\right.\nonumber\\
&+&\left.\left.2\tau_A^2+2\tau_B^2+\tau_A^4+\tau_B^4+20y^2\tau_A^2\tau_B^2-2\tau_A^4\tau_B^2\right.\right.\nonumber\\
&-&\left.\left.2\tau_A^2\tau_B^4+\tau_A^4\tau_B^4\right)\right.\nonumber\\
&-&\left.32I_{11}y\left(2y^2-1-\tau_A^2-\tau_B^2+\tau_A^2\tau_B^2\right)\right.\nonumber\\
&+&\left.J^A_{00}\left[15y^2-3+2\tau_A^2-2\tau_B^2-44y^2\tau_A^2+12\tau_A^2\tau_B^2\right.\right.\nonumber\\
&+&\left.\left.5\tau_A^4-11y^2\tau_A^4-2\tau_A^4\tau_B^2\right]\right.\nonumber\\
&+&\left.J^B_{00}\left[15y^2-3+2\tau_B^2-2\tau_A^2-44y^2\tau_B^2+12\tau_B^2\tau_A^2\right.\right.\nonumber\\
&+&\left.\left.5\tau_B^4-11y^2\tau_B^4-2\tau_B^4\tau_A^2\right]\right.\nonumber\\
&+&\left.{2\over 15}+{4\over 15}y^2+{34\over 3}y^2-2-5\tau_A^2-5\tau_B^2\right.\nonumber\\
&+&\left.11y^2\tau_A^2+11y^2\tau_B^2+2\tau_A^2\tau_B^2\right]\label{eq:zx}
\end{eqnarray}

As before, we can now proceed in two ways.

\subsection{Massive Particle Gravitational Scattering}

\begin{itemize}
\item[a)] {\bf Direct Evaluation:} In the massive case we work in the limit $y=1+\ldots$, whereby Eq. (\ref{eq:zx})
 becomes, using also $\tau_A,\,\tau_B<<1$,
\begin{eqnarray}
&&{\rm Disc}\,{\rm Amp}^{grav}_2(q)\stackrel{t<<m_A^2m_B^2}{\longrightarrow}-i{\kappa^4m_Am_B\over 1024\pi}\left[4I_{00}-32I_{11}+12J^A_{00}+12J^B_{00}+{28\over 3}+{2\over 5}\right]\nonumber\\
&=&-i\left[4\cdot \left(-{1\over 3}+i2\pi{m_Am_Bm_r\over p_0t}\right)-32\cdot \left(-1+i\pi{m_r\over 2p_0}\right)\right.\nonumber\\
&+&\left.12\left({\pi m_A\over \sqrt{t}}-1\right)+12\left({\pi m_B\over \sqrt{t}}-1\right)+{146\over 15}\right]\nonumber\\
&=&-i{\kappa^4m_Am_B\over 512\pi}\left[{41\over 5}+{6\pi(m_A+m_B)\over \sqrt{t}}+i4\pi{m_Am_Bm_r\over p_0t}L(1-{2t\over m_Am_B})\right]
\end{eqnarray}
We have then
\begin{equation}
{\rm Amp}^{grav}_2(q)=-{\kappa^4m_Am_B\over 1024\pi^2}\left[{41\over 5}L-6S(m_A+m_B)+i4\pi{m_Am_Bm_r\over p_0t}L+\ldots\right]\label{eq:mj}
\end{equation}

\item[b)] {\bf Feynman Integral Technique:}  Alternatively, we can write
\begin{eqnarray}
&&{\rm Disc}\,{\rm Amp}^{grav}_2(q)\stackrel{t<<m_A^2m_B^2}{\longrightarrow}i{\kappa^4m_Am_B\over 1024\pi}{\rm Disc}\,\left[
64\pi m_A^2m_B^2(I_4(s,t)+I_4(u,t))\right.\nonumber\\
&+&\left.128\pi tm_Am_B(I_4(s,t)-I_4(u,t))+96\pi m_A^2(I_3(p_1,q,m_A)+I_3(p_2,q,m_A))\right.\nonumber\\
&+&\left.96\pi m_B^2(I_3(-p_3,q,m_A)+I_3(-p_4,q,m_A))-8\pi{146\over 15}I_2(q)\right]
\end{eqnarray}
so again
\begin{eqnarray}
{\rm Amp}^{grav}_2(q)&=&{\kappa^4m_Am_B\over 1024\pi^2}\left[L({2\over 3}-16+6+6-{73\over 15})+6S(m_A+m_B)\right.\nonumber\\
&+&\left.i4\pi{L\over t}\sqrt{m_Am_B\over s-s_0}+\ldots\right]\nonumber\\
&=&-{\kappa^4m_Am_B\over 1024\pi^2}\left[{41\over 5}L-6S(m_A+m_B)+i4\pi{m_Am_Bm_r\over p_0t}L+\ldots\right]\nonumber\\
\quad\label{eq:nj}
\end{eqnarray}
which is identical to Eq. (\ref{eq:mj}).
\end{itemize}

The presence of the imaginary piece in Eqs. (\ref{eq:mj}) and (\ref{eq:nj}) is, of course, the gravitational phase shift and must be subtracted as in the electromagnetic case in order to define a proper second order potential.  Subtracting the second order Born amplitude
  \begin{eqnarray}
B_2^{grav}(q)&=&i\int{d^3\ell\over (2\pi)^3}{{\kappa^2\over 8}m_A^2\over |\boldsymbol{p}_f-\boldsymbol{\ell}|^2+\lambda^2}
{i\over {p_0^2\over 2m_r}-{\ell^2\over 2m_r}+i\epsilon}{{\kappa^2\over 8}m_B^2\over |\boldsymbol{\ell}-\boldsymbol{p}_i|^2+\lambda^2}\nonumber\\
&=&-i{\kappa^4\over 128\pi}m_A^2m_B^2{m_r\over p_0t}L\,,
\end{eqnarray}
the result is a well-defined second order gravitational potential
\begin{equation}
V_2^{grav}(r)=-{3G^24m_Am_B(m_A+m_B)\over r^2}-{41G^2m_Am_B\hbar\over 10\pi r^3}\label{eq:jf}
\end{equation}
The results found in Eqs. (\ref{eq:mj}) and (\ref{eq:nj}) as well as the potential given in Eq. (\ref{eq:jf}) are identical to those obtained using Feynman diagram methods in \cite{Khr03} and \cite{Bje03}.  However, they are obtained here with considerably less effort. The usual diagrammatic approach is a daunting one, due, among other things, to the necessity to include
\begin{itemize}
\item[i)] the sixth-rank tensor triple graviton coupling in numerous diagrams
\item[ii)] proper statistical and combinatorial factors for each diagram
\item[iii)] ghost contributions.
\end{itemize}
The challenge presented by this task is made clear by the fact since the seminal work of Donoghue in 1994~\cite{Don94} until the 2003 results obtained in \cite{Khr03} and \cite{Bje03}, there were a number of published calculations containing numerical errors~\cite{Don94},\cite{Muz95},\cite{Ham95},\cite{Akh97}.  The simplification  provided by the on-shell method described here is due essentially to the interchange of the order of integration and summation.  That is, in the conventional Feynman technique, one evaluates separate (four-dimensional) Feynman integrals for each separate diagram, which are then summed.  In the on-shell method, one first sums over the Compton scattering diagrams to obtain helicity amplitudes and {\it then} performs a (two-dimensional) intermediate state integration over solid-angle. There are a number of reasons why the latter procedure is more efficient.  For one, by using the explicitly gauge-invariant gravitational Compton amplitudes, the decomposition into separate and gauge-dependent diagrams is avoided.  Secondly, the various statistical/combinatorial factors are included automatically. Thirdly, because the intermediate states are on-shell, there is no ghost contribution~\cite{Pes95}.  Finally, the use of gravitational Compton amplitudes allows the use of factorization, which ameliorates the need to include the triple graviton coupling~\cite{Cho95},\cite{Ber02}.  The superposition of all these effects allows a relatively simple and highly efficient algebraic calculation of the gravitational scattering amplitude.

\subsection{Massive-Massless Particle Gravitational Scattering}

We saw in the previous section how on-shell methods combined with factorization to produce a relatively straightforward and efficient calculation of the second order gravitational scattering amplitude for two massive spinless particles $A$ and $B$.  Recently, similar techniques were used to study the problem of the gravitational scattering of massive and massless spin-zero particles, as a model of light bending around the sun~\cite{Bje15}, so it is interesting to adapt this formalism to handle this situation.  As in the electromagnetic case we take $\tau_A\stackrel{m_A\rightarrow 0}{\longrightarrow}i$, but $\tau_B<<1$ is unchanged.  Also, writing $s-m_B^2=2m_BE$, where E is the laboratory frame energy of the incident spinless particle, we shall again work in the small-angle scattering approximation $E>>\sqrt{t}$ so that $y=i|y|$ with $|y|=2E/\sqrt{t}>>1$.  Then

\begin{itemize}

\item[a)] {\bf Direct Evaluation:}  Setting $\tau_A=i$ Eq. (\ref{eq:zx}) becomes
\begin{eqnarray}
&&{\rm Disc}\,{\rm Amp}^{0-grav}_2(q)=-i{\kappa^4tm_B\xi_B^2\over 4096\pi E}\left[4I^{(0)}_{00}(4y^4-24y^2\tau_B^2+4\tau_B^4)\right.\nonumber\\
&-&\left.64I^{(0)}_{11}y(y^2-\tau_B^2)+J^{A(0)}_{00}(48y^2-16\tau_B^2)+J^B_{00}(15y^2-1\right.\nonumber\\
&-&\left.44y^2\tau_B^2-11y^2\tau_B^4-10\tau_B^2+7\tau_B^4)+{3\over 5}y^2+{47\over 15}+11y^2\tau_B^2-7\tau_B^2\right]\nonumber\\
\quad
\end{eqnarray}
Using $\tau_B<<1$ and making the small-angle scattering approximation, we find then
\begin{eqnarray}
{\rm Disc}\,{\rm Amp}^{0-grav}_2(q)&&\stackrel{|y|>>1}{\longrightarrow}-i{\kappa^4tm_B\over 4096\pi E}\left[16y^4I^{(0)}_{00}-64y^3I^{(0)}_{11}+48y^2J^{A(0)}_{00}+15y^2J^B_{00}
+{3\over 5}y^2\right]\nonumber\\
\quad\label{eq:xc}
\end{eqnarray}
Writing $y\simeq i{s-m_B^2+{t\over 2}\over m_B\sqrt{t}}$,
\begin{eqnarray}
&&{\rm Disc}\,{\rm Amp}^{0-grav}_2(q)\simeq -i{1\over 256\pi Em_B}\kappa^4(s+{t\over 2}-m_B^2)^2\nonumber\\
&\times&\left\{{1\over m_B^2t}(s+{t\over 2}-m_B^2)^2I^{(0)}_{00}\right.\nonumber\\
&+&\left.{4i\over m_B\sqrt{t}}(s+{t\over 2}-m_B^2)I^{(0)}_{11}-3J^{A(0)}_{00}-{15\over 16}J^B_{00}-{3\over 80}\right\}\nonumber\\
&=&-i{1\over 256\pi Em_B}\kappa^4(s+{t\over 2}-m_B^2)^2\nonumber\\
&\times&\left\{{(s+{t\over 2}-m_B^2)^2\over t}\left({1\over s-m_B^2}\ln{s-m_B^2\over m_B^2}+{1\over u-m_B^2}\ln{m_B^2\over u-m_B^2}\right)\right.\nonumber\\
&-&\left.2(s+{t\over 2}-m_B^2)\left({1\over s-m_B^2}\ln{s-m_B^2\over m_B^2}-{1\over u-m_B^2}\ln{m_B^2\over u-m_B^2}\right)\right.\nonumber\\
&+&\left.{3\over 2}\ln{-t\over \lambda^2}-{15\over 16}\left({\pi m_B\over \sqrt{t}}-1\right)-{3\over 80}\right\}
\end{eqnarray}
so, using
\begin{equation}
(s-m_B^2)^4,\,(u-m_B^2)^4=(s+{t\over 2}-m_B^2\mp{t\over 2})^4\simeq (s+{t\over 2}-m_B^2)^4\mp 2t(s+{t\over 2}-m_B^2)
\end{equation}
we find
\begin{eqnarray}
{\rm Amp}_2^{0-grav}(q)&\simeq&-{1\over 2048\pi^2 m_BE}\kappa^4(s-m_B^2)^2\left\{{1\over t}\left[(s-m_B^2)\ln{s-m_B^2\over m_B^2}\right.\right.\nonumber\\
&+&\left.\left.(u-m_B^2)\ln{u-m_B^2\over m_B^2}\right]L+3L^2+{15\over 4}\left(L+Sm_B\right)-{3\over 20}L\right\}\nonumber\\
&=&-{m_BE\over 512\pi^2}\kappa^4\left\{{1\over t}\left[2m_BE\ln\left({2E\over m_B}\right)\right.\right.\nonumber\\
&-&\left.\left.(2m_BE+t)\ln-\left({2E\over m_B}+{t\over m_B^2}\right)\right]L+3L^2+{15\over 4}\left(L+Sm_B\right)-{3\over 20}L\right\}\nonumber\\
\quad\label{eq:kf}
\end{eqnarray}

\item[b)] {\bf Feynman Integral Technique:} Alternatively we can write
\begin{eqnarray}
{\rm Disc}\,{\rm Amp}^{0-grav}_2(q)&\simeq &{i\over 64\pi m_BE}\kappa^4(s+{t\over 2}-m_B^2)^2{\rm Disc}\,{1\over t}\nonumber\\
&\times&\left\{(s-m_B^2+{t\over 2})^2(I^{(0)}_4(s,t)+I^{(0)}_4(u,t))\right.\nonumber\\
&+&\left.2(s+{t\over 2}-m_B^2)(I^{(0)}_4(s,t)-I^{(0)}_4(u,t))\right.\nonumber\\
&-&\left.{3\over 2}t(I_3^{(0)}(p_1,q,0)+I_3^{(0)}(p_2,q,0))\right.\nonumber\\
&+&\left.{15\over 8}m_B^2(I_3(-p_3,q,m_B)+I_3(-p_4,q,m_B))\right.\nonumber\\
&-&\left.{3\over 40}I_2(q)\right\}
\end{eqnarray}
so
\begin{eqnarray}
{\rm Amp}_2^{0-grav}(q)&\simeq&-{1\over 2048\pi^2m_BE}\kappa^4(s-m_B^2)^2\left\{{1\over t}\left[(s-m_B^2)\ln{s-m_B^2\over m_B^2}\right.\right.\nonumber\\
&+&\left.\left.(u-m_B^2)\ln{u-m_B^2\over m_B^2}\right]L+3L^2+{15\over 4}\left(L+Sm_B\right)-{3\over 20}L\right\}\nonumber\\
\quad
\quad\label{eq:lf}
\end{eqnarray}
\end{itemize}

Again the results found in Eqs. (\ref{eq:kf}) and (\ref{eq:lf}) are identical and are in agreement with the result calculated in \cite{Bje16}, but are obtained via a more efficient algebraic method.  Defining
\begin{eqnarray}
{\rm Amp}_2^{0-grav}(q)&\equiv&-{\kappa^4\over 4m_BE}\left[bo^\phi(s,t)I_4^{(0)}(s,t)+bo^\phi(u,t)I_4^{(0)}(u,t)+{1\over 2}t_{12}^\phi(I_3^{(0)}(p_1,q,0)\right.\nonumber\\
&+&\left.I_3^{(0)}(p_2,q,0))+{1\over 2}t_{34}^\phi(I_3(-p_3,q,m_B)+I_3(-p_4,q,m_B))+bu^\phi I_2(q)\right]\nonumber\\
\quad
\end{eqnarray}
and using the result that $u-m_B^2\stackrel{t\rightarrow 0}{\longrightarrow} -(s-m_B^2)$, we find
\begin{equation}
{bo^\phi(s,t)\over s-m_B^2}+{bo^\phi(u,t)\over u-m_B^2}={1\over 32}\left[(s-m_B^2)^3+(u-m_B^2)^3\right]\stackrel{t\rightarrow 0}{\longrightarrow}-{3t(s-m_B^2)^2\over 32}=-t_{12}^\phi
\end{equation}
so that the BCJ relation is satisfied and again serves as a confirmation of our calculation~\cite{Ber08}.  In addition, we note that the sum of the two terms in the top line of Eq. (\ref{eq:lf}) becomes imaginary, corresponding to a gravitational scattering phase, which must be subtracted.  In terms of $E$, the laboratory frame energy of the massless particle, the effective gravitational potential is then
\begin{eqnarray}
V_2^{0-grav}(r)&=&-\int{d^3q\over (2\pi)^3}e^{-i\boldsymbol{q}\cdot\boldsymbol{r}}\left({\rm Amp}_2^{0-grav}(q)-B_2^{0-grav}(q)\right)\,,\nonumber\\
&=&{15\over 4}{G^2m_B^2E\over r^2}-{15-{3\over 5}\over 4\pi}{Gm_BE\hbar\over r^3}-{12G^2m_B\hbar\over \pi r^3}\ln{r\over r_0}
\end{eqnarray}
and agrees with the form given in \cite{Bje15}.  (Note that though \cite{Bje15} also used on-shell methods, the results were obtained diagram by diagram.)

\section{Conclusion}

We have shown above that the use of on-shell techniques accompanied by the use of factorization in the gravitational case has produced a straightforward and efficient way to evaluate higher order electromagnetic and/or gravitational scattering, both in the scattering of spinless particles with $m_A,\,m_B\neq 0$ and in the case $m_A=0,\,m_B\neq 0$. Results found in this way were shown to agree exactly with those obtained by more cumbersome Feymman diagram techniques.  In the case of the electromagnetic interaction, this can be seen since the usual diagrammatic approach involves evaluation of the individual (and gauge-dependent) contributions from the bubble, triangle, box and cross-box diagrams already shown in Figure 3. This simplification is much more significant in the case of gravitational scattering, since the Feynman diagram calculation involves
\begin{itemize}
\item[a)] not only the bubble, triangle, box, and cross-box diagrams considered in the electromagnetic case and shown in Figure 3 (but now with tensor vertices associated with the energy-momentum tensor replacing the vector vertices associated with the electromagnetic current and fourth-rank tensor graviton propagators replacing second rank tensor photon propagators), but also

\item[b)] completely new 5a),5b) vertex-bubble and 5c),5d) vertex-triangle diagrams involving the sixth-rank-tensor triple graviton vertex, together with the vacuum polarization diagram, which involves {\it two} triple graviton vertices, as shown in Figure 5e.  In addition, the vacuum polarization contribution must be modified by the ghost loop diagram shown in Figure 5f).
\end{itemize}

\begin{figure}
\begin{center}
\epsfig{file=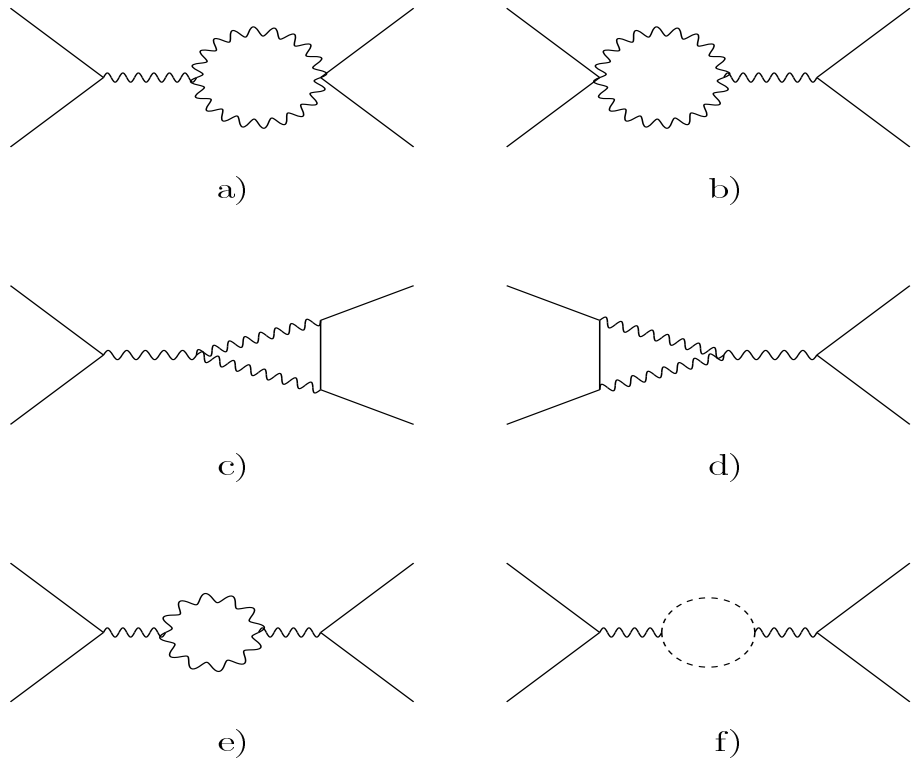,height=10cm,width=12cm}
\caption{{Shown are the a),b) vertex-bubble, c),d) vertex-triangle, e) vacuum polarization and f) ghost-loop diagrams contributing to spinless particle gravitational scattering, all but the ghost-loop involving the triple-graviton vertex. Here the solid lines represent the massive spinless particles, while the wiggly lines represent gravitons.}}
\end{center}
\end{figure}
As mentioned above, similar methods are used by Bjerrum-Bohr et al.\cite{Bje16} to obtain these results.  The difference between the procedure used therein and and that described above is that in \cite{Bje16} the calculation is performed covariantly, with the low energy limit taken only at the end.  The technique used above, involving taking the low energy limit before the integration is performed, allows a much simpler path to the desired results.

The simplification produced by use of these methods should allow straightforward extension to the situation that either or both particles carry spin or if both are massless.  These situations are presently under study.

\begin{center}
{\bf\large Appendix A:  Solid-Angle Averaged Integrals $I_{mn}$}
\end{center}

The fundamental angular-averaged quantities $J^i_{00},\,I_{00},\,$ and $I_{11}$ have been given by Feinberg and Sucher~\cite{Fei88}.  In the case of $J^i_{00},\,\, i=A,B$
\begin{equation}
J^i_{00}=<{1\over d_i}>={1\over 4\pi}\int_0^{2\pi}d\phi\int_{-1}^1 {dx_i\over \tau_i^2+x_i^2}={1\over \tau_i}{\rm tan}^{-1}{1\over \tau_i}={1\over \tau_i}\left({\pi\over 2}-{\rm tan}^{-1}\tau_i\right)
\end{equation}
while
\begin{eqnarray}
I_{00}&=&<{1\over d_Ad_B}>={1\over 2\tau_A\tau_B}(F_++{\pi\over N_+})\nonumber\\
I_{11}&=&<{x_Ax_B\over d_Ad_B}>={1\over 2}(F_-+{\pi\over N_+})
\end{eqnarray}
where
\begin{equation}
F_\pm(s,t)=\pm{1\over N_-}{\rm tan}^{-1}{N_-\over D_+}-{1\over N_+}{\rm tan}^{-1}{N_+\over D_-}
\end{equation}
with
\begin{eqnarray}
N_\pm(s,t)&=&(\tau_A^2+\tau_B^2+1-y^2\pm 2\tau_A\tau_By)^{1\over 2}\nonumber\\
D_\pm(s,t)&=&y\pm\tau_A\tau_B
\end{eqnarray}

 By repeated use of the algebraic identities $x_i^2=d_i-\tau_i^2,\,\,i=A,B$ one can write all other $I_{mn}$ in terms of "seed" quantities---$I_{00},\,I_{11},\,J^A_{00},\,J^B_{00},\,<1>=1$:
\begin{eqnarray}
I_{20}&=&J^B_{00}-\tau_A^2I_{00}\nonumber\\
I_{02}&=&J^A_{00}-\tau_B^2I_{00}\nonumber\\
I_{22}&=&1-\tau_B^2J^B_{00}-\tau_A^2J^A_{00}+\tau_A^2\tau_B^2I_{00}\nonumber\\
I_{31}&=&y(1-\tau_B^2J^B_{00})-\tau_A^2I_{11}\nonumber\\
I_{13}&=&y(1-\tau_A^2J^A_{00})-\tau_B^2I_{11}\nonumber\\
I_{40}&=&-{1\over 2}(1-3y^2)+{1\over 2}\left((1-y^2)+\tau_B^2(1-3y^2)-2\tau_A^2\right)J^B_{00}+\tau_A^4I_{00}\nonumber\\
I_{04}&=&-{1\over 2}(1-3y^2)+{1\over 2}\left((1-y^2)+\tau_A^2(1-3y^2)-2\tau_B^2\right)J^A_{00}+\tau_B^4I_{00}\nonumber\\
I_{42}&=&{1\over 3}-\tau_A^2+{1\over 2}\tau_B^2(1-3y^2)-{1\over 2}\tau_B^2\left((1-y^2)+\tau_B^2(1-3y^2)-2\tau_A^2\right)J^B_{00}\nonumber\\
&+&\tau_A^4J^A_{00}-\tau_A^4\tau_B^2I_{00}\nonumber\\
I_{24}&=&{1\over 3}-\tau_B^2+{1\over 2}\tau_A^2(1-3y^2)-{1\over 2}\tau_A^2\left((1-y^2)+\tau_A^2(1-3y^2)-2\tau_B^2\right)J^A_{00}\nonumber\\
&+&\tau_B^4J^B_{00}-\tau_A^2\tau_B^4I_{00}\nonumber\\
I_{33}&=&y\left({1\over 3}-\tau_A^2-\tau_B^2\right)+y\tau_A^4J^A_{00}+y\tau_B^4J^B_{00}+\tau_A^2\tau_B^2I_{11}\nonumber\\
I_{44}&=&{1\over 5}+{2\over 15}y^2+{1\over 2}\tau_A^2(1-3y^2)+{1\over 2}\tau_B^2(1-3y^2)+\tau_A^2\tau_B^2\nonumber\\
&+&-{1\over 2}\tau_A^2\left((1-y^2)+\tau_A^2(1-3y^2)+\tau_A^2\tau_B^2\right)J^A_{00}\nonumber\\
&+&-{1\over 2}\tau_B^2\left((1-y^2)+\tau_B^2(1-3y^2)+\tau_A^2\tau_B^2\right)J^B_{00}\nonumber\\
&+&\tau_A^4\tau_B^4I_{00}
\end{eqnarray}

\begin{center}
{\bf\large Appendix B: Connection with Feynman Scalar Integrals}
\end{center}

It is useful to relate the fundamental "seed" quantities $I_{00},\,I_{11},J^A_{00},\,J^B_{00},\,<1>$
to familiar Feynman scalar integrals.

\begin{itemize}
\item[a)] {\bf Scalar Box+Cross-Box Diagram}:  Noting that
\begin{equation}
{1\over p_1\cdot\ell p_2\cdot\ell}={2\over t}\left({1\over p_1\cdot\ell}+{1\over p_2\cdot\ell}\right)\quad{\rm and}\quad {1\over p_3\cdot\ell p_4\cdot\ell}={2\over t}\left({1\over p_3\cdot\ell}+{1\over p_4\cdot\ell}\right)
\end{equation}
and defining the on-shell scalar box and cross-box diagrams via
\begin{eqnarray}
{\rm Amp}_{box}&\equiv& I_4(s,t)=\int{d^4\ell\over (2\pi)^4}{1\over \ell^2(\ell-q)^2}\left({1\over ((\ell-p_1)^2-m_A^2)((\ell+p_3)^2-m_B^2)}\right)\nonumber\\
&=&\int{d^4\ell\over (2\pi)^4}{1\over \ell^2(\ell-q)^2}\left({1\over ((\ell-p_2)^2-m_A^2)((\ell+p_4)^2-m_B^2)}\right)\nonumber\\
{\rm Amp}_{c-box}&\equiv& I_4(u,t)=\int{d^4\ell\over (2\pi)^4}{1\over \ell^2(\ell-q)^2}\left({1\over ((\ell-p_1)^2-m_A^2)((\ell+p_4)^2-m_B^2)}\right)\nonumber\\
&=&\int{d^4\ell\over (2\pi)^4}{1\over \ell^2(\ell-q)^2}\left({1\over ((\ell-p_2)^2-m_A^2)((\ell-p_3)^2+m_B^2)}\right)
\end{eqnarray}
we have, then,
\begin{equation}
\int{d^4\ell\over (2\pi)^4}{1\over \ell^2(\ell-q)^2p_1\cdot\ell p_2\cdot\ell p_3\cdot\ell p_4\cdot\ell}=-{32\over t^2}(I_4(s,t)+I_4(u,t))
\end{equation}
Using the Feinberg-Sucher continuation
\begin{equation}
{1\over p_1\cdot\ell p_2\cdot\ell p_3\cdot\ell p_4\cdot\ell}\stackrel{FS}{\longrightarrow} {16\over t^2m_A^2\xi_A^2m_B^2\xi_B^2d_Ad_B}
\end{equation}
so that
\begin{equation}
{\rm Disc}\,(I_4(s,t)+I_4(u,t))=-{1\over 16\pi m_A^2\xi_A^2m_B^2\xi_B^2}<{1\over d_Ad_B}>
\end{equation}
or
\begin{equation}
I_{00}=<{1\over d_Ad_B}>=-16\pi m_A^2\xi_A^2m_B^2\xi_B^2\,{\rm Disc}\,(I_4(s,t)+I_4(u,t))\label{eq:kx}
\end{equation}

\item[b)] {\bf Box-Cross-Box Diagram}:
Similarly, defining $P_A=p_1-p_2$ and $P_B=p_3-p_4$,
\begin{equation}
{P_A\cdot\ell P_B\cdot\ell\over p_1\cdot\ell p_2\cdot\ell p_3\cdot\ell p_4\cdot \ell}=\left({1\over p_1\cdot\ell}-{1\over p_2\cdot\ell}\right)\left({1\over p_3\cdot\ell}-{1\over p_4\cdot\ell}\right)
\end{equation}
so
\begin{equation}
\int{d^4\ell\over (2\pi)^4}{P_A\cdot\ell P_B\cdot\ell\over \ell^2(\ell-q)^2p_1\cdot\ell p_2\cdot\ell p_3\cdot\ell p_4\cdot\ell}=-8(I_4(s,t)-I_4(u,t))
\end{equation}
Then, since
\begin{equation}
{P_A\cdot\ell P_B\cdot\ell\over p_1\cdot\ell p_2\cdot\ell p_3\cdot\ell p_4\cdot\ell}\stackrel{FS}{\longrightarrow}{-16x_Ax_B\over tm_A\xi_Am_B\xi_Bd_Ad_B}
\end{equation}
we have
\begin{equation}
{\rm Disc}\,(I_4(s,t)-I_4(u,t))={1\over 4\pi tm_A\xi_Am_B\xi_B}I_{11}
\end{equation}
or
\begin{equation}
I_{11}=4\pi tm_A\xi_Am_B\xi_B\,{\rm Disc}\,(I_4(s,t)-I_4,(u,t))
\end{equation}

\item[c)] {\bf Triangle Diagram}:
Defining
\begin{equation}
I_3(p,q,m)\equiv\int{d^4\ell\over (2\pi)^4}{1\over \ell^2(\ell-q)^2((\ell-p)^2-m^2)}
\end{equation}
we note
\begin{equation}
{P_B\cdot\ell P_B\cdot\ell\over p_1\cdot\ell p_2\cdot\ell p_3\cdot\ell p_4\cdot \ell}=\left[4+{t\over 2}\left({1\over p_3\cdot\ell}+{1\over p_4\cdot\ell}\right)\right]{2\over t}\left({1\over p_1\cdot\ell}+{1\over p_2\cdot\ell}\right)
\end{equation}
Then since
\begin{equation}
{P_B\cdot\ell P_B\cdot\ell\over p_1\cdot\ell p_2\cdot\ell p_3\cdot\ell p_4\cdot \ell}\stackrel{FS}{\longrightarrow}{-16x_B^2\over tm_A^2\xi_A^2d_Ad_B}
\end{equation}
we have
\begin{eqnarray}
&&{\rm Disc}\,\left[-8(I_4(s,t)+I_4(u,t))+{16\over t}(I_3(p_1,q,m_A)+I_3(p_2,q,m_A))\right]\nonumber\\
&=&-{2\over \pi tm_A^2\xi_A^2}I_{02}
\end{eqnarray}
Using
\begin{equation}
I_{02}=J^A_{00}-\tau_B^2I_{00}
\end{equation}
and Eq. (\ref{eq:kx}) we find
\begin{equation}
{\rm Disc}\,(I_3(p_1,q,m_A)+I_3(p_2,q,m_A))=-{1\over 8\pi m_A^2\xi_A^2J^A_{00}}
\end{equation}
or
\begin{equation}
J^A_{00}=-8\pi m_A^2\xi_A^2\,{\rm Disc}\,(I_3(p_1,q,m_A)+I_3(p_2,q,m_A))
\end{equation}
Similarly,
\begin{equation}
J^B_{00}=-8\pi m_B^2\xi_B^2\,{\rm Disc}\,(I_3(-p_3,q,m_B)+I_3(-p_4,q,m_B))
\end{equation}

\item[d)] Finally, and trivially
\begin{equation}
<1>=8\pi\,{\rm Disc}\,I_2(q)
\end{equation}
\end{itemize}

\begin{center}
{\bf\large Appendix C:  Integral Evaluation}
\end{center}

\noindent{\bf Massive Case:}

\begin{itemize}
\item[a):]\, In the case of the direct evaluation, we require the seed forms
\begin{eqnarray}
I_{00}&=&{1\over 2\tau_A\tau_B}\left(F_++{\pi\over N_+}\right)\nonumber\\
I_{11}&=&{1\over 2}\left(F_-+{\pi\over N_+}\right)
\end{eqnarray}
where
\begin{equation}
F_\pm(s,t)=\pm{1\over N_-}{\rm tan}^{-1}{N_-\over D_+}-{1\over N_+}{\rm tan}^{-1}{N_+\over D_-}
\end{equation}
with
\begin{eqnarray}
N_\pm(s,t)&=&(\tau_A^2+\tau_B^2+1-y^2\pm 2\tau_A\tau_By)^{1\over 2}\nonumber\\
D_\pm(s,t)&=&y\pm\tau_A\tau_B
\end{eqnarray}
and
\begin{equation}
y(s_0,t)={2s_0+t-2m_A^2-2m_B^2\over 4m_A\xi_Am_B\xi_B}\stackrel{t\rightarrow 0}{\longrightarrow}1+{\cal O}(t)
\end{equation}
Then
\begin{equation}
N_\pm(s,t)= \left((1+\tau_A^2)(1+\tau_B^2)-(y\mp \tau_A\tau_B)^2\right)^{1\over 2}= \left((1+\tau_A^2)(1+\tau_B^2)-D_\mp^2\right)^{1\over 2}
\end{equation}
and
\begin{equation}
D_\pm={f(s,t)+t\pm t\over 4m_A\xi_Am_B\xi_B}
\end{equation}
where
\begin{equation}
f(s,t)=4m_Am_B+2(s-s_0)
\end{equation}
Also
\begin{equation}
(1+\tau_A^2)(1+\tau_B^2)=(1+{t\over 4m_A^2\xi_A^2})(1+{t\over 4m_B^2\xi_B^2})={1\over \xi_A^2\xi_B^2}
\end{equation}
so
\begin{eqnarray}
{N_\pm^2\over D_\mp^2}&=&-1+{16m_A^2m_B^2\over (f(s,t)+t\mp t)^2}=-1+{1\over (1+{s-s_0\over 2m_Am_B}+{t\mp t\over 4m_Am_B})^2}\nonumber\\
&=&-1+{1\over \left(1+{2sp^2\over m_Am_B(s-(m_A-m_B)^2)}+{t\mp t\over 4m_Am_B}\right)^2}\nonumber\\
&=&-{4sp^2\over m_Am_B(s-(m_A-m_B)^2)}-{t\mp t\over 2m_Am_B}+\ldots\nonumber\\
&\simeq&-{p^2\over m_r^2}-{t\mp t\over 2m_Am_B}
\end{eqnarray}
which agrees with the exact relations
\begin{eqnarray}
N_+&=&-i{p\sqrt{s}\over m_A\xi_Am_B\xi_B}\nonumber\\
N_-&=&-i{\sqrt{p^2s+m_A\xi_Am_B\xi_Bt}\over m_A\xi_Am_B\xi_B}
\end{eqnarray}
Thus
\begin{equation}
{1\over 3}\left[\left({N_+\over D_-}\right)^2-\left({N_-\over D_+}\right)^2\right]={t\over 3m_Am_B}+\ldots
\end{equation}
and
\begin{eqnarray}
{1\over D_+}-{1\over D_-}&=&{D_--D_+\over D_+D_-}={-8tm_A\xi_Am_B\xi_B\over f(s,t)(f(s,t)+2t)}={-t\over 2m_Am_B}+\ldots\nonumber\\
-{1\over D_+}-{1\over D_-}&=&-2+\ldots
\end{eqnarray}
Then
\begin{eqnarray}
F_+(s,t)&=&\left({1\over D_+}-{1\over D_-}-{1\over 3}{N_-^2\over D_+^3}-{1\over 3}{N_+^2\over D_-^3}+\ldots\right)\nonumber\\
&=&-{t\over 2m_Am_B}+{t\over 3m_Am_B}+\ldots=-{t\over 6m_Am_B}+\ldots\nonumber\\
F_-(s,t)&=&-{1\over D_+}-{1\over D_-}
\end{eqnarray}
We have then
\begin{eqnarray}
I_{00}&=&<{1\over d_Ad_B}>={1\over 2\tau_A\tau_B}(F_+(s,t)+{\pi\over N_+})={2m_Am_B\over t}\left(-{t\over 6m_Am_B}+i{\pi m_r\over p}+\ldots\right)\nonumber\\
&=&-{1\over 3}+2i\pi {m_rm_Am_B\over pt}+\ldots\nonumber\\
I_{11}&=&<{x_Ax_B\over d_Ad_B}>={1\over 2}(F_-(s,t)+{\pi\over N_+})={1\over 2}\left(-2+i\pi{m_r\over p}+\ldots\right)\nonumber\\
&=&-1+i\pi{m_r\over 2p}+\ldots
\end{eqnarray}

\item[b):]\, In the case of the Feynman diagrams we need the seed forms $I_4(s,t)$ and $I_4(u,t)$.  We use
\begin{eqnarray}
I_4(s,t)&=&\int{d^4\ell\over (2\pi)^4}{1\over (\ell^2-\lambda^2)((\ell-q)^2-\lambda^2)((\ell-p_1)^2-m_A^2)((\ell-p_3)^2-m_B^2)}\nonumber\\
&=&{i\over 8\pi^2t\sqrt{s^2-2s(m_A^2+m_B^2)+(m_A^2-m_B^2)^2}}\ln{-t\over \lambda^2}\nonumber\\
&\times&\ln{s-(m_A+m_B)^2-\sqrt{s^2-2s(m_A^2+m_B^2)+(m_A^2-m_B^2)^2}\over s-(m_A+m_B)^2+\sqrt{s^2-2s(m_A^2+m_B^2)+(m_A^2-m_B^2)^2}}\nonumber\\
\quad\label{eq:vw}
\end{eqnarray}
where $s=(p_1-p_3)^2,\,u=(p_1-p_4)^2$ and $t=2p_1\cdot (p_3+p_4)=q^2$.  Write $s=(m_A+m_B)^2+s-s_0$ so
$$s^2-2s(m_A^2+m_B^2)+(m_A^2-m_B^2)^2=4m_Am_B(s-s_0)+(s-s_0)^2$$
Because of the logarithmic and square root cuts, the integrals must be carefully defined.  The correct choices for the box and cross-box integrals
can be identified from the results for $I_{00},\,I_{11}$---
\begin{eqnarray}
I_4(s,t)&+&I_4(u,t)\nonumber\\
&=&-{iL\over 16\pi^2m_Am_Bt}\left({p_1\cdot(p_3+p_4)\over 3m_Am_B}-4\pi i\sqrt{m_Am_B\over s-s_0}+\ldots\right)\nonumber\\
&=&{iL\over 32\pi^2m_A^2m_B^2}\left(-{1\over 3}-{4i\pi m_Am_B\over t}\sqrt{m_Am_B\over s-s_0}+\ldots\right)\nonumber\\
I_4(s,t)&-&I_4(u,t)\nonumber\\
&=&{iL\over 8\pi^2m_Am_Bt}\left(1-i\pi \sqrt{m_Am_B\over s-s_0}+\ldots\right)
\end{eqnarray}
\end{itemize}

\noindent {\bf Massless Case:}

We begin with the case of direct evaluation.

\begin{itemize}
\item[a):]  In order to evaluate the modified triangle integral, we introduce a cutoff $\lambda$ via $\tau_A=i(1-2{\lambda^2\over t})$
\begin{equation}
J^A_{00}={1\over 2}\int_{-1}^1 {dx_A\over \tau_A^2+x_A^2}\stackrel{m_A\rightarrow 0}{\longrightarrow}-{1\over 2}\ln{-t\over \lambda^2}
\end{equation}
In order to define box and cross box integrals we then note
\begin{equation}
D_\pm=y\pm\tau_A\tau_B\stackrel{m_A\rightarrow 0}{\longrightarrow}i\left({s-m_B^2\over 2\tau_Bm_B^2\xi_B^2}+\tau_B\pm \tau_B\right)=i{u-m_B^2\over 2\tau_Bm_B^2\xi_B^2},i{s-m_B^2\over 2\tau_Bm_B^2\xi_B^2}
\end{equation}
and
\begin{eqnarray}
{N_\pm^2\over D_\mp^2}&=&(1+\tau_A^2)(1+\tau_B^2)-1=-1+4{\lambda^2\over t}(1+\tau_B^2)\left({u,s-m_B^2\over 2\tau_Bm_B^2\xi_B^2}\right)^{-2}\nonumber\\
&=&i^2\left[1-2{\lambda^2\over t}(1+\tau_B^2)\left({u,s-m_B^2\over 2\tau_Bm_B^2\xi_B^2}\right)^{-2}\right]^2
\end{eqnarray}
so that we need ${\rm tanh}^{-1}{N_\pm\over D_\mp}$.  Since
\begin{eqnarray}
&&{\rm tanh}^{-1}\left[1-{2{\lambda^2\over t}(1+\tau_B^2)\over \left({u,s-m_B^2\over 2\tau_Bm_B^2\xi_B^2}\right)^2}\right]=-{1\over 2}\ln{{\lambda^2\over t}(1+\tau_B^2)\over \left({u,s-m_B^2\over 2\tau_Bm_B^2\xi_B^2}\right)^2}\nonumber\\
&=&\left[\ln{{u,s}-m_B^2\over m_B^2\xi_B^2}-{1\over 2}\ln{4\lambda^2\over t}\tau_B^2(1+\tau_B^2)\right]\nonumber\\
&=&\left[\ln{{u,s}-m_B^2\over m_B^2\xi_B^2}-{1\over 2}\ln{\lambda^2\over m_B^2\xi_B^2}(1+\tau_B^2)\right]
\end{eqnarray}
Then
\begin{eqnarray}
I^{(0)}_{11}&\simeq &{1\over 2}F_-(s,t)=-{1\over 2}\left[{1\over N_-}{\rm tan}^{-1}{N_-\over D_+}+{1\over N_+}{\rm tan}^{-1}{N_+\over D_-}\right]\nonumber\\
&=&-{i\over 2}\left[{1\over D_+}{\rm tan}^{-1}{N_-\over D_+}+{1\over D_-}{\rm tan}^{-1}{N_+\over D_-}\right]\nonumber\\
&=&{i\over 2}\sqrt{t}m_B\xi_B\left[{1\over s-m_B^2}\ln{s-m_B^2\over m_B^2\xi_B^2}-{1\over u-m_B^2}\ln{u-m_B^2\over m_B^2\xi_B^2}+\ldots\right]\nonumber\\
\quad
\end{eqnarray}
and
\begin{eqnarray}
I^{(0)}_{00}&\simeq &{1\over 2\tau_A\tau_B}F_+(s,t)={1\over 2\tau_A\tau_B}\left[{1\over N_-}{\rm tan}^{-1}{N_-\over D_+}-{1\over N_+}{\rm tan}^{-1}{N_+\over D_-}\right]\nonumber\\
&=&{m_B\xi_B\over \sqrt{t}}\left[{1\over D_+}{\rm tan}^{-1}{N_-\over D_+}-{1\over D_-}{\rm tan}^{-1}{N_+\over D_-}\right]\nonumber\\
&=& m_B^2\xi_B^2\left[{1\over s-m_B^2}\ln{s-m_B^2\over m_B^2\xi_B^2}+{1\over u-m_B^2}\ln{u-m_B^2\over m_B^2\xi_B^2}+\ldots\right]\nonumber\\
\quad
\end{eqnarray}

In the case of the Feynman diagram technique

\item[b):]
Using Eq. (\ref{eq:vw}), we have
\begin{eqnarray}
I_4(s,t)^{(0)}&=&{i\over 16\pi^2(s-m_B^2)t}\left(\ln {s-m_B^2\over \mu^2}+\ln{s-m_B^2\over m_B^2}\right)L\nonumber\\
I_4(u,t)^{(0)}&=&{i\over 16\pi^2(u-m_B^2)t}\left(\ln {u-m_B^2\over \mu^2}+\ln{u-m_B^2\over m_B^2}\right)L\nonumber\\
\end{eqnarray}
where $\mu^2$ is a regulator mass.  Similarly
\begin{equation}
I_3(p_1,q,0)={i\over 32\pi^2t}L^2
\end{equation}
\end{itemize}

\begin{center}
{\bf\large Acknowledgement}
\end{center}

It is a pleasure to acknowledge numerous helpful conversations with John Donoghue, which served to greatly clarify the material discussed above. This work is supported in part by the National Science Foundation under award NSF PHY11-25915.

\end{document}